\def\asca{{\it ASCA\/}}

\def\chandra{{\it Chandra\/}}

\def\einstein{{\it Einstein\/}}

\def\rosat{{\it ROSAT\/}}

\def\uhuru{{\it Uhuru\/}}
\def\xeus{{\it XEUS\/}}
\def\xmm{{\it XMM-Newton\/}}

\def\ltsima{$\; \buildrel < \over \sim \;$}
\def\simlt{\lower.5ex\hbox{\ltsima}}
\def\gtsima{$\; \buildrel > \over \sim \;$}
\def\simgt{\lower.5ex\hbox{\gtsima}}
        
\documentclass[preprint]{aastex}

\begin{document}


\title{The Chandra Deep Field North Survey. V. 1~Ms Source Catalogs}


\author{W.N.~Brandt,$^1$ 
D.M.~Alexander,$^1$ 
A.E.~Hornschemeier,$^1$ 
G.P.~Garmire,$^1$ 
D.P.~Schneider,$^1$
A.J.~Barger,$^{2,3,4}$ 
F.E.~Bauer,$^1$ 
P.S.~Broos,$^1$ 
L.L.~Cowie,$^2$ 
L.K.~Townsley,$^1$
D.N.~Burrows,$^1$ 
G.~Chartas,$^1$
E.D.~Feigelson,$^1$  
R.E.~Griffiths,$^5$ 
J.A.~Nousek,$^1$ and 
W.L.W.~Sargent$^6$ 
}

\footnotetext[1]{Department of Astronomy \& Astrophysics, 525 Davey Laboratory, 
The Pennsylvania State University, University Park, PA 16802}

\footnotetext[2]{Institute for Astronomy, University of Hawaii, 
2680 Woodlawn Drive, Honolulu, HI 96822} 

\footnotetext[3]{Department of Astronomy, University of Wisconsin-Madison,
475 N. Charter Street, Madison, WI 53706}

\footnotetext[4]{Hubble Fellow and Chandra Fellow at Large}

\footnotetext[5]{Department of Physics, Carnegie Mellon University, Pittsburgh, PA 15213}

\footnotetext[6]{Palomar Observatory, California Institute of Technology, Pasadena, CA 91125}


\begin{abstract}
An extremely deep X-ray survey ($\approx 1$~Ms) of the Hubble Deep Field North and its environs
($\approx 450$~arcmin$^2$) has been performed with the Advanced CCD Imaging Spectrometer
on board the {\it Chandra X-ray Observatory\/}. This is one of the two deepest X-ray
surveys ever performed; for point sources near the aim point it 
reaches 0.5--2.0~keV and 2--8~keV flux limits of 
$\approx 3\times 10^{-17}$~erg~cm$^{-2}$~s$^{-1}$ and
$\approx 2\times 10^{-16}$~erg~cm$^{-2}$~s$^{-1}$, respectively. 
Here we provide source catalogs along with details of the observations, 
data reduction, and technical analysis. Observing conditions, such as 
background, were excellent for almost all of the exposure. 

We have detected 370 distinct point sources: 
360 in the 0.5--8.0~keV band, 
325 in the 0.5--2.0~keV band, 
265 in the 2--8~keV band, and 
145 in the 4--8~keV band. 
Two new \chandra\ sources in the HDF-N itself are reported and discussed. 
Source positions are accurate to within 0.6--1.7$^{\prime\prime}$
(at $\approx 90$\% confidence) depending 
mainly on the off-axis angle. We also detect two 
highly significant extended X-ray sources and several other 
likely extended X-ray sources. 

We present basic number count results for sources located near the 
center of the field. Source densities of 
$7100^{+1100}_{-940}$~deg$^{-2}$ (at \hbox{$4.2\times 10^{-17}$~erg~cm$^{-2}$~s$^{-1}$}) and 
$4200^{+670}_{-580}$~deg$^{-2}$ (at \hbox{$3.8\times 10^{-16}$~erg~cm$^{-2}$~s$^{-1}$})
are observed in the soft and hard bands, respectively.  
\end{abstract}




\section{Introduction}

We have performed a deep X-ray survey ($\approx 1$~Ms) of the Hubble Deep Field
North (\hbox{HDF-N}; Williams et~al. 1996, hereafter W96; 
Ferguson, Dickinson, \& Williams 2000)
and its environs with the {\it Chandra X-ray Observatory\/} (hereafter \chandra):
the \chandra\ Deep Field North (hereafter CDF-N) survey.
This is one of the two deepest X-ray surveys ever conducted, the other being the 
\chandra\ Deep Field South survey (e.g., Tozzi et~al. 2001; P.~Rosati et~al., 
in preparation). Near the aim point, the CDF-N observation 
reaches 0.5--2.0~keV and 2--8~keV limiting fluxes of 
$\approx 3\times 10^{-17}$~erg~cm$^{-2}$~s$^{-1}$ and
$\approx 2\times 10^{-16}$~erg~cm$^{-2}$~s$^{-1}$, respectively;
these flux limits are $\approx 40$ and $\approx 400$ times fainter
than achieved by pre-\chandra\ missions. 
In Figure~1 we compare the 0.5--2.0~keV flux limit and solid angle 
of this survey to those of several other extragalactic X-ray surveys.
This survey has both the sensitivity 
and positional accuracy needed to complement the deepest surveys at other 
wavelengths, and much of the $\approx 450$~arcmin$^2$ area surveyed has extensive
radio, submillimeter, infrared, and optical coverage (see
Livio, Fall, \& Madau 1998 and Ferguson et~al. 2000 for reviews). 
 
The main goals of the CDF-N survey are
(1) to understand the broad-band emission and nature of the sources 
producing the X-ray background in the 0.5--8.0~keV band, and
(2) to investigate the X-ray emission properties of physically interesting
sources identified at other wavelengths.
Thus far, we have presented results for the HDF-N 
(Hornschemeier et~al. 2000, hereafter Paper~I; Brandt et~al. 2001a, Paper~IV)
as well as for larger fields centered on the HDF-N
(Hornschemeier et~al. 2001, Paper~II; 
Garmire et~al. 2001, Paper~III; Barger et~al. 2001).
Due to the fact that the \chandra\ observations were performed in an incremental
manner from 1999 November to 2001 March, the studies above were all performed
with less than the full $\approx 1$~Ms exposure.

In this paper, we provide source catalogs derived from the $\approx 1$~Ms data
set along with details of the observations, data reduction, and technical analysis. Our 
intention is to provide this information to the community in as timely a manner as 
possible. We have deliberately avoided follow-up investigations and detailed scientific 
interpretation in this paper; these will be presented in subsequent papers 
(and many such results have been presented in the papers
cited above). For example, companion papers by 
Alexander et~al. (2001, Paper~VI), 
Brandt et~al. (2001b, Paper~VII), 
A.J. Barger et~al., in preparation, and 
F.E. Bauer et~al., in preparation 
present results on 
optically faint X-ray sources, 
X-ray emission from Lyman break galaxies, 
optical follow-up imaging and spectroscopy, and
extended X-ray sources.  

In \S2 we describe the observations and data reduction. In \S3 we describe the
data analysis and results, with emphasis on the detection of point sources (\S3.2)
and extended sources (\S3.3). We also present basic number count results for
point sources in \S3.2. Our conclusions and summary are presented in \S4.

The Galactic column density along this line of sight
is $(1.6\pm 0.4)\times 10^{20}$~cm$^{-2}$ (Stark et~al. 1992).
$H_0=70$~km~s$^{-1}$ Mpc$^{-1}$, 
$\Omega_{\rm M}=1/3$, and 
$\Omega_{\Lambda}=2/3$
are adopted throughout this paper. 
Coordinates throughout this paper are J2000.
The HDF-N itself is centered at
$\alpha_{2000}=$~12$^{\rm h}$ 36$^{\rm m}$ 49\fs 4,
$\delta_{2000}=$~$+62^\circ$12$^{\prime}$58$^{\prime\prime}$, 
corresponding to 
$l=125\fdg 888$, $b=54\fdg 828$ (W96). 


\section{Observations and Data Reduction}

\subsection{Instrumentation}

All observations of the CDF-N field were performed with the 
Advanced CCD Imaging Spectrometer (ACIS; G.P. Garmire et~al., in 
preparation) on \chandra\ 
(Weisskopf et~al. 2000).\footnote{For additional information on 
the ACIS and \chandra\ see the \chandra\ Proposers' Observatory Guide 
at http://asc.harvard.edu/udocs/docs.} ACIS consists 
of ten CCDs designed for efficient X-ray detection and 
spectroscopy. Four of the CCDs (ACIS-I; CCDs I0--I3) are arranged in 
a $2\times 2$ array with each CCD tipped slightly to approximate the 
curved focal surface of the \chandra\ High Resolution Mirror Assembly 
(HRMA). The remaining six CCDs (ACIS-S; CCDs S0--S5) are set in a 
linear array and are tipped to approximate the Rowland circle of 
the objective gratings that can be inserted behind the HRMA. The 
CCD which lies on-axis in ACIS-I is I3, and the full ACIS-I field of 
view is $16\farcm 9\times 16\farcm 9$.

The ACIS pixel size is $\approx 0\farcs 492$. The 95\%
encircled-energy radius for an energy of 1.5~keV at the aim point
is $\approx 1\farcs 8$. This radius increases to 
$\approx 2\farcs 7$ when $4^{\prime}$ from the aim point and
$\approx 7\farcs 5$ when $8^{\prime}$ from the aim point
(Feigelson, Broos, \& Gaffney 2000; Jerius et~al. 2000; 
M.~Karovska and P.~Zhao 2001, 
private communication).\footnote{Feigelson et~al. (2000) is available 
at http://www.astro.psu.edu/xray/acis/memos/memoindex.html.}
Note that at higher energies the aim-point PSF is significantly
broader than at 1.5~keV. 

\subsection{Observational Parameters and Conditions}

The CDF-N was observed in twelve separate observations as detailed 
in Table~1. The total exposure time was 975.3~ks. The HDF-N was 
placed near the aim point of the ACIS-I array during all observations, 
and care was taken to keep the HDF-N away from the gaps between the CCDs. 
The focal-plane temperature, which governs several characteristics of the 
CCD behavior, in particular the Charge Transfer Inefficiency (CTI), was 
$-110^\circ$\,C during the first three observations and $-120^\circ$\,C 
during the others. 
When observing with ACIS-I, two CCDs from ACIS-S, typically S2 and
S3, can be operated. CCD S3 was turned off during the HDF-N observations 
due to the higher background level of this device; this property could 
cause telemetry saturation during background flares. CCD S2, however, 
was operated.

The region covered by the \chandra\ observations is considerably larger 
than $16\farcm 9\times 16\farcm 9$ due to the different observation
pointings and roll angles (see Table~1). These variations were necessary 
to satisfy the roll constraints of \chandra\ while keeping the HDF-N 
itself near the aim point and away from the gaps between the CCDs. 
The average aim point, weighted by exposure time, is
$\alpha_{2000}=$~12$^{\rm h}$ 36$^{\rm m}$ $48\fs 1$,
$\delta_{2000}=$~$+62^\circ$13$^{\prime}$53$^{\prime\prime}$. 
The aim points of the individual observations are separated from
the average aim point by $1\farcm 6$--$3\farcm 1$; most are 
within $2\farcm 1$ of the average aim point. 
Due to the large off-axis angle of CCD S2 during these observations, it 
has low sensitivity. Therefore, we only include data from S2 
when these data overlap part of the sky also covered by ACIS-I in other 
observations. 

Background light curves have been inspected for all of the observations. 
All but one are free from strong flaring due to ``space weather'' and are 
stable to within $\approx 20$\%. The only observation with substantial 
flaring is 2344; during $\approx 30$~ks of this observation the background 
was $\approx 2$ times higher than nominal. However, the data quality 
during the flaring was still sufficiently high to provide useful
scientific information so these data were not excluded; the
exclusion of these data would have little impact on the analysis 
or results presented here. 

\subsection{Data Reduction}

The versions of the \chandra\ X-ray Center (hereafter CXC)  
pipeline software used for basic processing of the data are listed in Table~1. 
In the reduction and analysis below, \chandra\ Interactive Analysis of 
Observations ({\sc ciao}) Version~2 tools were used whenever 
possible.\footnote{See http://asc.harvard.edu/ciao/.} 
Tools for ACIS Real-time Analysis ({\sc tara}; 
Broos et~al. 2000)
and custom software were also used.\footnote{{\sc tara} is 
available at http://www.astro.psu.edu/xray/docs.}

All data were corrected for the radiation damage sustained by the CCDs during 
the first few months of \chandra\ operations using the procedure of 
Townsley et~al. (2000).\footnote{The software associated with the correction method of
Townsley et~al. (2000) is available at http://www.astro.psu.edu/users/townsley/cti/.}
This procedure partially corrects for the positionally dependent 
grade distribution due to inefficient charge transfer in the radiation-damaged
CCDs. It also partially corrects for quantum efficiency losses, which are most 
significant in data acquired at $-110^{\circ}$\,C (see Townsley et~al. 2000 
and Paper~II for discussion of the remaining small quantum efficiency 
losses incurred).

We have removed bad columns, bad pixels, and cosmic ray afterglows 
as flagged by the CXC using 
the ``status'' information in the event files.\footnote{Cosmic ray 
afterglows occur in front-side illuminated ACIS CCDs when 
charge from an incident cosmic ray is released slowly, causing a 
series of spurious events in the same CCD pixel over several 
sequential frames. These events can resemble a real cosmic X-ray 
source. See http://cxc.harvard.edu/ciao/advanced\_documents.html
for information on cosmic ray afterglows.}
We have only used data taken during times within the CXC-generated good-time 
intervals. The standard pixel randomization was removed as part
of the aspect correction procedure described in \S3.1.\footnote{See 
http://asc.harvard.edu/cal/Hrma/hrma/misc/oac/dd\_psf/dd\_randomiz.html.}


\section{Data Analysis and Results}

\subsection{Image and Exposure Map Creation}

In this paper we report on the emission detected in four 
standard X-ray bands: 
0.5--8.0~keV (full band), 
0.5--2.0~keV (soft band),
2--8~keV (hard band) and 
4--8~keV (ultrahard band). 
We have adopted 8~keV (rather than the often-used 10~keV) 
as the full-band, hard-band, and ultrahard-band 
maximum energy because from 8--10~keV the 
effective area of the HRMA is steeply decreasing with energy while 
the background is increasing.\footnote{For further information on the 
HRMA performance and \chandra\ background see Chapters~4 and 6 of
the \chandra\ Proposers' Observatory Guide at 
http://asc.harvard.edu/udocs/docs. Also see the memos on \chandra\ background 
at http://asc.harvard.edu/cal/Links/Acis/acis/WWWacis\_cal.html.}
Inspection and searching of the data 
revealed no significant sources in the 8--10~keV band.
We have employed the two grade sets defined in Table~2. As described in
Paper~IV, the use of the ``restricted ACIS grade set'' in addition
to the ``standard \asca\ grade set'' improves our ability to detect faint
sources in some cases. All photometry below, however, is reported using
the standard \asca\ grade set. 

We have registered the data sets following \S3.1 of Paper~II. 
Briefly, we registered all the data sets to the 
coordinate frame of observation 966 using 11--17 bright X-ray 
sources detected in the individual observations within $\approx 6\arcmin$ 
of the aim point; registration is accurate to within $\approx 0\farcs 4$. 
Absolute X-ray source positions were obtained by matching 
72 sources from the registered total data set to 1.4~GHz radio 
sources detected by Richards (2000); these 1.4~GHz sources have 
accurate ($\simlt 0\farcs 3$) positions, and the 1.4~GHz coverage 
encompasses the entire \chandra\ field. Comparison with these
sources allowed us to remove shift, rotation, and plate-scale
effects. \chandra\ positions used in the data set registration 
and absolute astrometry determination were found using the 
wavelet-based source detection algorithm {\sc wavdetect} 
(Dobrzycki et~al. 1999; Freeman et~al. 2001) following \S3.2.1. 

Figure~2 shows the accuracy of our astrometric solution by matching
the full-band \chandra\ sources presented in \S3.2.1 with 1.4~GHz 
sources from Richards (2000). There are 241 1.4~GHz sources within 
the \chandra\ field. In the matching, we consider only the 74 
\chandra\ sources that match with 1.4~GHz sources to within 
$2\farcs 5$ (the 72 sources used to determine the absolute astrometry
in the previous paragraph excluded the two outliers in Figure~2 with
offsets $>1\farcs 5$). The vast majority of these 74 matches are 
expected to be correct, but $\approx 2$ are expected to be false 
matches. We also note that in some cases the X-ray 
source may be offset from the radio source even though both are 
associated with the same galaxy (e.g., a galaxy with a radio-emitting
nuclear starburst that also has an off-nuclear ``super-Eddington'' X-ray 
binary). Figure~2 shows that the X-ray positions are usually good 
to within $\approx 0\farcs 6$ for off-axis angles $<5^{\prime}$. 
At larger off-axis angles, where the HRMA point spread function (PSF)
rapidly broadens and becomes complex, the positional accuracy, as 
expected, degrades. Most sources have positions good to within 
$1\arcsec$, but positional offsets up to $\approx$~1\farcs 5--2\farcs 0 
are possible. We do not find any systematic errors in our astrometric 
solution larger than $\approx 0\farcs 25$ in size. 

In Figures~3 and 4 we show raw and adaptively smoothed images of our
field in each of the four standard X-ray bands. These images have been made 
using the standard \asca\ grade set.
The adaptively smoothed images have been corrected for spatial variations
of the effective exposure time using the exposure maps described below. 
The exposure maps were adaptively smoothed using the same ``scale maps'' 
as for the images themselves, and we excluded regions where the 
adaptively smoothed effective exposure time was less than 50~ks. 
The adaptively smoothed images were not used for source detection, but 
they do show many of the detected X-ray sources more clearly than the 
raw images. Note that some of the sources discussed below are not 
visible in the adaptively smoothed images; these sources fall below 
the significance level of the adaptive smoothing used to make the
images. 
In Figure~5 we show a color composite of the adaptively smoothed
soft-band, hard-band and ultrahard-band images. Soft sources appear
red, moderately hard sources appear green, and the hardest sources
appear blue. 
In Figure~6 we show an adaptively smoothed full-band image of the HDF-N 
itself and its immediate environs; this image will be discussed further 
in \S3.2.3. 

We have made maps of effective exposure time, defined as the 
equivalent amount of exposure time for a source located at the 
aim point, following the basic procedure described in \S3.2 of Paper~II
(see Figure~7). These ``exposure maps'' take into account the effects 
of vignetting, gaps between the CCDs, bad column filtering, and 
bad pixel filtering. One exposure map has been created for each
of the standard bands, and the maps were sampled every fourth pixel
in both right ascension and declination. In creating these, we have 
assumed a typical power-law spectrum with a photon index 
of $\Gamma=1.4$ (this is the slope of the X-ray background in 
the \chandra\ band).  
In Figure~8 we show a cumulative plot of the survey solid angle 
as a function of effective exposure in the full band. 

\subsection{Point-Source Detection and Results}

\subsubsection{Point-Source Detection and Parameterization}

Point-source detection in each standard band was performed with {\sc wavdetect}. 
{\sc wavdetect} was run using a ``$\sqrt{2}$~sequence'' of wavelet scales; 
scales of 1, $\sqrt{2}$, 2, $2\sqrt{2}$, 4, $4\sqrt{2}$, and 8 pixels were 
used. We found this choice of wavelet scales to give the best overall 
performance across the field, but we will also discuss the results for larger 
wavelet scales in \S3.2.2. In the {\sc wavdetect} source detection the average
aim point defined in \S2.2 was used when calculating off-axis angles. 
Our key criterion for source detection is that a source must be found with a 
false-positive probability threshold of $1\times 10^{-7}$ in at least one 
of the four standard bands using either the standard \asca\ or restricted 
ACIS grade sets. We have {\it not\/} run {\sc wavdetect} with a low 
false-positive probability threshold and then performed {\it ex post facto\/} 
processing of the detected sources to attempt to separate true from spurious 
detections; such a process would be subjective and counter to the 
philosophy of {\sc wavdetect} (Freeman et~al. 2001; 
P.E. Freeman 2001, private communication). 
Our detection criterion is fairly conservative; fainter but real sources
are undoubtedly present in the field. However, our work on the CDF-N 
data thus far shows that we have struck an appropriate balance between 
sensitivity and source reliability. 

Conservatively treating the eight images searched as entirely independent, 
$\approx 5$ false sources are expected statistically for the case of a 
uniform background. In reality, the background is far from uniform due
to the large variation of effective 
exposure time across the field (see Figure~7). In
addition, the background increases near bright point sources 
due to the PSF ``wings.'' It is difficult to quantify precisely the effects 
of a non-uniform background upon the performance of {\sc wavdetect}. However,
based upon the amount of area where background gradients are present
and inspection of the sources detected in this area, we would
not expect the number of false sources to be increased by more than a
factor of $\sim$~2--3. Less than $\approx 4$\% of the sources discussed 
below should be false.

Extensive analysis of the {\sc wavdetect} source positions revealed that
a significant fraction of the sources at off-axis angles $\simgt 4^{\prime}$ 
suffered from centroiding errors of $0.2$--$2^{\prime\prime}$. These 
apparently arise as a result of both limitations in the {\sc wavdetect} 
centroiding method as well as the fact that there is not a single \chandra\ 
PSF applicable for any given source (due to the different observation 
pointings and roll angles described in \S2.2, the spatial profile of
any given point source is the superposition of several \chandra\ 
PSFs).\footnote{Some information on 
this issue is given as Case~\#4460 of the CXC Helpdesk at 
http://cxc.harvard.edu/helpdesk/.} Testing and consultation with the CXC 
showed that the {\sc wavdetect} centroiding could be significantly
improved by running {\sc wavdetect} without information about the
\chandra\ PSF (P.E.~Freeman and D.E.~Harris 2001, private communication). 
Therefore, in as many cases as possible, we have replaced the original
{\sc wavdetect} source positions with those from {\sc wavdetect} runs
made excluding PSF information; note that these ``no-PSF'' runs were only
used for positional replacement and not for source detection. Positional
replacement was performed for 86\% of the detected sources; for
the remaining sources it was not possible due to either a non-detection
or a multiple detection in the the no-PSF run. The average 
positional improvement was $\approx 0\farcs 3$ but in some
cases was as large as $\approx 1\farcs 5$ (as determined by matching
with Richards 2000 sources at 1.4~GHz). 

{\sc wavdetect} was also used to search the standard-band images for lower-significance, 
cross-band counterparts to the highly significant sources already detected at the 
$1\times 10^{-7}$ level in at least one of the four bands; in these runs we used 
a false-positive probability threshold of $1\times 10^{-5}$. We found 110 
cross-band counterparts in this manner. Since the spatial-matching requirement 
greatly reduces the number of pixels being searched, the statistically expected 
number of false cross-band matches is small ($\simlt 0.5$). 

All of the standard-band source lists created in the source detection 
described above were merged to create the point-source catalog given as Table~3. 
For cross-band matching, a matching radius of $\le 2\farcs 5$ was used 
for sources within $6\arcmin$ of the average aim point. For larger
off-axis angles, a matching radius of $\le 4\farcs 0$ was used. 
These matching radii were chosen based on inspection of histograms 
showing the number of matches obtained as a function of angular 
separation (see \S2 of Boller et~al. 1998); with these radii 
the mismatch probability is $\approx 1$\% over the entire field. 

Manual correction of the {\sc wavdetect} results was required in 
several cases. For example, {\sc wavdetect} detections 
of the brightest components of highly extended sources were 
removed since these sources will be discussed separately in \S3.3. In 
four cases, we have removed sources whose centroids appear to 
lie outside the field of view; {\sc wavdetect} had only detected
the PSF wings of these sources, and their positions and
count rates are not well defined. Manual separation of a
few close doubles was required, and we have manually determined 
the position of each separated source. These sources suffer 
larger photometric errors due to the difficulty of the 
separation process. It was also necessary to perform manual 
photometry for two sources near the edge of the field of view 
and a few faint sources located close to much brighter sources. 
We have flagged all sources requiring manual correction
in Column~16 of Table~3 (see below).  

Below we explain the columns in Table~3. 


\begin{itemize}

\item
Column~1 gives the source number. Sources are listed in order of right ascension. 

\item
Columns~2 and 3 give the right ascension and declination, respectively. These positions 
have been determined by {\sc wavdetect} when possible.
Whenever possible, we quote the position determined in the full band; when a source 
is not detected in the full band we use, in order of priority, the soft-band position, 
hard-band position, or ultrahard-band position. In addition, we adopt a soft-band
position from a no-PSF run over a full-band position when full-band positional 
replacement from a no-PSF run was not possible (see above). The priority ordering of
position choices above was designed to generally maximize the signal-to-noise
of the data being used for positional determination. To avoid truncation 
error, we quote the positions to higher precision than in the International 
Astronomical Union approved names beginning with the 
acronym ``CXO~HDFN.''\footnote{See http://cxc.harvard.edu/udocs/naming.html.}

\item
Column~4 gives the positional error. Sources within $5\arcmin$ of the average
aim point have positional errors of $0\farcs 6$. Sources farther than $5\arcmin$ 
from the average aim point have positional errors given by the empirically 
determined equation: 

$$\Delta=0.6+\left({\theta-5\arcmin\over 8.75\arcmin}\right)~{\rm arcsec}~~~~~({\rm for}~\theta>5\arcmin)$$

\noindent
where $\Delta$ is the positional error in arcsec and 
$\theta$ is the off-axis angle in arcmin (compare with Figure~2). 
The positional error does not appear to be a strong function of the
number of source counts (although a mild dependence is probably
present); this is largely due to the relatively sharp core of the \chandra\ PSF. 
The stated positional errors are for $\approx 90$\% confidence, and the 
accuracy of our astrometric solution is discussed in \S3.1.

\item
Columns~5--8 give the counts in the four standard bands; here and hereafter 
``FB'' indicates full band, 
``SB'' indicates soft band, 
``HB'' indicates hard band, and 
``UHB'' indicates ultrahard band.
All values are for the standard \asca\ grade set, and they have not been 
corrected for vignetting. Source counts and $1\sigma$ statistical errors 
(from Gehrels 1986) have been calculated using circular aperture photometry; 
extensive testing showed this method to be more reliable than the 
{\sc wavdetect} photometry. The circular aperture was centered at the
position given in columns 2 and 3 for all bands. A source-free local 
background has been subtracted, and unexposed regions were masked. 

For sources with fewer than 1000 full-band counts, we have chosen the aperture  
radii based on the encircled-energy function of the \chandra\ PSF as determined
using the CXC's {\sc mkpsf} software (Feigelson et~al. 2000; 
Jerius et~al. 2000; M.~Karovska and P.~Zhao 2001, private 
communication). In the soft band where the image quality is the best, the 
aperture radius was set to the 95\% encircled-energy radius of the PSF, and in 
the other bands the 90\% encircled-energy radius of the PSF was used. Appropriate 
aperture corrections were applied to the source counts. 

For sources with more than 1000 full-band counts, systematic errors in the 
aperture corrections often exceed the expected errors from photon statistics 
when the apertures described in the previous paragraph are used. Therefore, 
for such sources we used larger apertures to minimize the importance
of the aperture corrections; this is appropriate since these bright sources 
dominate over the background. We set the aperture radii to be twice those
used in the previous paragraph and manually inspected these sources to 
verify that the measurements were not contaminated by neighboring objects. 

We have performed several consistency checks to verify the quality of
the photometry. For example, we have checked that the sum of the counts
measured in the soft and hard bands does not differ from the counts measured 
in the full band by an amount larger than that expected from measurement 
error. Systematic errors in our photometry are estimated to be $\simlt 4$\%. 
Photometry more accurate than this will require improved modeling of the 
\chandra\ PSF at large off-axis angles by the CXC (M.~Karovska and P.~Zhao 2001, 
private communication) as well as detailed treatment of the fact that there is 
not a single \chandra\ PSF applicable for any given source (due to the
different observation pointings and roll angles described in \S2.2). 
In addition, it will probably be necessary to make the aperture 
corrections for each source dependent upon its spectral shape. 

We have verified that the cosmic ray afterglow removal procedure (see \S2.3) 
has not led to significant systematic photometric errors. Due to the low count
rates of our sources, incident X-ray photons are almost never incorrectly 
flagged as afterglow events. 

When a source is not detected in a given band, an upper limit is calculated. 
All upper limits are determined using circular apertures as above. 
When the number of counts in the aperture is $\leq 10$, the upper limit is 
calculated using the Bayesian method of Kraft, Burrows, \& Nousek (1991) for 
95\% confidence. The uniform prior used by these authors results in fairly 
conservative upper limits (see Bickel 1992), and other reasonable choices of 
priors do not materially change our scientific results. 
For larger numbers of counts in the aperture, upper limits are calculated
at the $3\sigma$ level for Gaussian statistics. 

\item
Column~9 gives the band ratio, defined as the ratio of counts between the hard and soft bands. 
Errors for this quantity are calculated following the ``numerical method'' described in 
\S1.7.3 of Lyons (1991); this avoids the failure of the standard approximate variance formula 
when the number of counts is small (see \S2.4.5 of Eadie et~al. 1971). Note that the
error distribution is not Gaussian when the number of counts is small. Quoted band ratios 
have been corrected for differential vignetting between the hard band and soft band using 
the appropriate exposure maps. 

\item
Column~10 gives the effective photon index ($\Gamma$) for a power-law model with the 
Galactic column density. This has been calculated based on the band ratio in 
column~9 whenever the number of counts is not ``low.'' A source with a low number 
of counts is defined as being
(1) detected in the soft band with $<30$ counts and not detected in the hard band, 
(2) detected in the hard band with $<15$ counts and not detected in the soft band, 
(3) detected in both the soft and hard bands, but with $<15$ counts in each, or
(4) detected only in the full band. 
When the number of counts is low, the photon index is poorly constrained 
and set to $\Gamma=1.4$, a representative value for faint sources
that should give reasonable fluxes.

\item
Column~11 gives the effective exposure time derived from the full-band exposure map 
(see \S3.1 for details of the exposure maps). 

\item
Columns~12--15 give observed-frame fluxes in the four standard bands; fluxes are in units of 
$10^{-15}$~erg~cm$^{-2}$~s$^{-1}$. They have been corrected for vignetting but are not 
corrected for absorption by the Galaxy or intrinsic to the source. For a power-law model 
with $\Gamma=1.4$, the soft-band and hard-band Galactic absorption corrections 
are $\approx 4.2$\% and $\approx 0.1$\%, respectively. 
Fluxes have been computed using 
the counts in columns~5--8, 
the appropriate exposure maps, and  
the spectral slopes given in column~10. 

\item
Column~16 gives notes on the sources. 
``H'' and ``C'' denote objects lying in the \hbox{HDF-N} and the Caltech area, 
respectively (see Figure~3). 
``O'' refers to objects that have large cross-band positional offsets ($>2\farcs 5$).  
``NP'' refers to objects where the source position could not be updated to that from a no-PSF run. 
``M'' refers to sources where the photometry was performed manually. 
``S'' refers to close double sources where manual separation was required. 
For further explanation of many of these notes, see the above text in this 
section on manual correction of the {\sc wavdetect} results.

\end{itemize}

\noindent
Some of the sources in Table~3 have been presented in Papers~I--IV. The
source properties in Table~3 supersede those presented in earlier papers. 

Figures~9--12 display the basic properties of the sources in Table~3. 
In Figure~9 we plot the positions of the sources detected in the
soft band and hard band; this format removes the illusory effect 
produced by the changing PSF size across the field of view. 
Figure~10 shows the distribution of full-band effective exposure
time; this plot includes the ten sources that were
not detected in the full band (all of the sources that were not detected
in the full band were detected in the soft band, and the full-band and
soft-band effective exposure times are similar). 
Figure~11 shows ``postage-stamp'' images in the full band for all 
detected sources. 
Figure~12 displays the band ratio as a function of soft-band count 
rate for the detected \chandra\ sources. This plot shows the same
qualitative behavior as those from shallower surveys 
(e.g., Paper~IV; Tozzi et~al. 2001) and populates the faint flux 
region more densely. The sources generally become harder 
at low soft-band count rates, although there is substantial
dispersion. At the lowest soft-band count rates, the source 
population is heterogeneous; highly absorbed AGN, high-redshift AGN, 
low-luminosity AGN, starburst galaxies, and normal galaxies all
appear to make significant contributions (e.g., Paper~II;
Paper~IV; Paper~VI; Tozzi et~al. 2001). Understanding in detail the
nature and relative contributions of the source populations represented
in Figure~12 will require extensive optical spectroscopic and
multiwavelength follow-up studies. 

We have inspected the sources in Table~3 for spatial extent, and they
are generally consistent with being pointlike (see Figure~11). 
The constraints on spatial extent are 20--40\% worse than for a 
single ACIS-I observation because, due to the different observation 
pointings and roll angles described in \S2.2, 
there is not a single \chandra\ PSF applicable for any given source. 
The constraints at off-axis angles of $\simlt 5^{\prime}$ are also
substantially tighter than those at larger off-axis angles. 
Spatially extended sources are discussed in \S3.3. 

In Table~4 we summarize the source detections in the four standard
bands, and in Table~5 we summarize the number of sources detected
in one band but not another. 

Our faintest soft-band sources have $\approx 6$ counts (about one 
every 1.8 days), and our faintest hard-band sources have 
$\approx 10$ counts; these sources are detected near the aim 
point. For a $\Gamma=1.4$ power-law model with the Galactic column 
density, the corresponding 0.5--2.0~keV and 2--8~keV flux limits are 
$\approx 2.9\times 10^{-17}$~erg~cm$^{-2}$~s$^{-1}$ and
$\approx 1.9\times 10^{-16}$~erg~cm$^{-2}$~s$^{-1}$, respectively. 
Of course, these flux limits vary across the field 
of view. 
Using the restricted ACIS grade set, the background level in our
region of highest exposure (see Figures~7 and 8) is
$3.6\times 10^{-2}$~count~pixel$^{-1}$ in the soft band and
$1.1\times 10^{-1}$~count~pixel$^{-1}$ in the hard band.
In both of these bands we are far from being background limited
for point source detection near the aim point.  

\subsubsection{Supplementary Point Sources at Large Off-Axis Angles}

Unfortunately, there is no single choice of wavelet scales in 
{\sc wavdetect} that provides perfect performance across an entire
\chandra\ field (Freeman et~al. 2001; P.E. Freeman 2001, private 
communication). Compromises are required due to, for example, the
dependence of the PSF size and shape on off-axis angle. In addition, due 
to the different observation pointings and roll angles described in 
\S2.2, there is not a single \chandra\ PSF applicable for any given 
source in this particular field. The wavelet scales used in \S3.2.1 have 
been empirically found to provide very good performance across most of
the field, but adding further large wavelet scales (e.g., $8\sqrt{2}$ 
and 16 pixels) can improve the detection effectiveness at large off-axis 
angles where the PSF is broad. However, the addition of these scales 
can cause real sources at smaller off-axis angles to be missed due 
to the incorrect merging of two sources. In addition, the ``rejection 
rule'' of {\sc wavdetect}, designed to suppress Poisson fluctuations on 
scales smaller than the PSF, has limitations that can cause incorrect 
source rejections when large wavelet scales are employed 
(see \S3.2.3 of Freeman et~al. 2001; P.E. Freeman 2001, private 
communication). For example, incorrect source rejections may occur
when only the relatively sharp core of the PSF is apparent above
the background level. 

To address this issue and provide the best possible performance at
large off-axis angles, we have created a supplementary catalog of
11 additional sources found by {\sc wavdetect} at large off-axis angles 
when wavelet scales of $8\sqrt{2}$ and 16 pixels are added to those 
used in \S3.2.1 (see Table~6). We have followed the same basic 
methodology as was used when making the main catalog (see \S3.2.1);
we have again adopted a false-positive probability threshold of 
$1\times 10^{-7}$ as the key criterion for source detection.
The catalog columns are the same as for the main catalog. 
Because the resulting sources tend to be the weakest 
ones detectable at the largest off-axis angles, their 
properties are in general less well defined than those of the 
sources in the main catalog. We have manually inspected these sources 
to confirm their reality. 

\subsubsection{Comparison with Previous Results for the HDF-N}

A comparison of the \chandra\ sources detected in the HDF-N 
here and in Paper~IV gives good agreement. All but one of the 12 sources
detected in Paper~IV are detected here as well. The one source not
detected here is CXOHDFN~J123643.9+621249; this faint source is known to
be variable (see Paper~IV) and has dropped in flux, leading to a reduced 
signal-to-noise ratio in the longer observation.\footnote{We note that
there may be other faint, variable sources like CXOHDFN~J123643.9+621249
that could be recovered by searching the data using many time segments. 
However, addressing this issue comprehensively is a challenging task 
and is beyond the scope of this paper.} There is still a 
notable positive fluctuation at the location 
of CXOHDFN~J123643.9+621249 (see Figure~6), and it is detected 
in the soft band if {\sc wavdetect} is run with a false-positive 
probability threshold of $1\times 10^{-6}$. One other small 
difference from Paper~IV is that the position of CXOHDFN~J123649.4+621347 
has moved closer to the nucleus of its $z=0.089$ elliptical host 
galaxy (see \S3.1 of Paper~IV); it is no longer clear that this faint 
X-ray source lies outside the nucleus. 

With the additional data presented here, two new HDF-N X-ray sources 
have been discovered: CXOHDFN~J123656.6+621245 and CXOHDFN~J123657.4+621210. 
Both sources appear to be associated with optically bright HDF-N galaxies, 
extending the trend noted in Paper~IV. The first is W96 source 3-610.1 at
$z=0.518$ (Cohen et~al. 2000), one of the optically brightest ($V_{606}=21.30$) 
spirals in the HDF-N. The second is W96 source 3-965.0 at 
$z=0.665$ (Cohen et~al. 2000), a bright ($V_{606}=22.23$) elliptical near 
the edge of the HDF-N. The full-band X-ray luminosities of these galaxies 
of $\approx 1\times 10^{41}$~erg~s$^{-1}$ and 
$\approx 3\times 10^{41}$~erg~s$^{-1}$, respectively, 
can be explained via either low-luminosity 
AGN or stellar remnants (e.g., X-ray binaries). 
Both of these galaxies were included in the stacking analysis of 
Paper~IV, and their individual detections with additional data 
support the validity of the stacking analysis. 

We note that there are several other positive fluctuations in the 
HDF-N visible in Figure~6. While some of these align with bright HDF-N 
galaxies (e.g., the positive fluctuation near 
$\alpha_{2000}=$~12$^{\rm h}$ 36$^{\rm m}$ 52\fs 8,
$\delta_{2000}=$~$+62^\circ$13$^{\prime}$54$^{\prime\prime}$ 
aligns with W96 source 2-736.0, a $z=1.355$ irregular) and may 
well be real, they are not formally detected according to the 
criteria in \S3.2.1 and will not be discussed further here. 

\subsubsection{Number Counts for the Point Sources}

We have calculated cumulative number counts, $N(>S)$, for the soft and 
hard bands using the data in Table~3 and the basic method described in 
Gioia et~al. (1990). The results are shown in Figure~13, along with 
some results from \rosat\ and \asca. The CDF-N number counts 
do not extend to higher fluxes due to the limited number of bright sources 
available in our data. 

We have been conservative in our source selection, taking several 
steps to prevent incompleteness (due to the varying sensitivity as a 
function of spatial position) from affecting the $N(>S)$ curves. 
To avoid incompleteness in the soft band, we have only used data which 
satisfy the soft-band requirements in Table~7 on 
(1) distance from the average aim point and 
(2) soft-band effective exposure time. 
The second requirement has the effect of removing areas affected by 
the gaps between the CCDs, where incompleteness is difficult to 
assess (see Figure~7). In Table~7 we also list the solid angle
satisfying the above requirements. We have only extended the 
soft-band $N(>S)$ down to $4.2\times 10^{-17}$~erg~cm$^{-2}$~s$^{-1}$ 
since, even near the average aim point, 
incompleteness may affect the $N(>S)$ at fainter fluxes; sources
below this flux typically have $\simlt 8$ soft-band counts. At this 
flux limit, the source density is $7100^{+1100}_{-940}$~deg$^{-2}$. 
The \chandra\ and \rosat\ constraints on the soft-band $N(>S)$ are
consistent in the limited region of overlap. We have fit the CDF-N 
$N(>S)$ curve in the flux range $5\times 10^{-17}$~erg~cm$^{-2}$~s$^{-1}$
to $2\times 10^{-15}$~erg~cm$^{-2}$~s$^{-1}$ using a maximum likelihood
technique (Murdoch, Crawford, \& Jauncey 1973). The best fit is

$$N(>S)=3970\left({S\over 1\times 10^{-16}}\right)^{-0.67\pm 0.14}.$$

To avoid incompleteness in the hard band, we have only used data 
which satisfy the hard-band requirements in Table~7. We have 
only extended the hard-band $N(>S)$ down to 
$3.8\times 10^{-16}$~erg~cm$^{-2}$~s$^{-1}$ since, even near the 
average aim point, incompleteness 
may affect the $N(>S)$ at fainter fluxes; sources
below this flux typically have $\simlt 12$ hard-band counts. At this 
flux limit, the source density is $4200^{+670}_{-580}$~deg$^{-2}$. 
There is very little overlap between the \chandra\ constraints on the 
hard-band $N(>S)$ and those from earlier missions. 
The best fit to the CDF-N $N(>S)$ curve in the flux range above
$1.5\times 10^{-15}$~erg~cm$^{-2}$~s$^{-1}$ is 

$$N(>S)=2820\left({S\over 1\times 10^{-15}}\right)^{-1.0\pm 0.3}.$$

\noindent
At fainter fluxes the number counts clearly flatten (see Figure~13b), 
and parameterization of them requires a detailed analysis of the
differential counts (see Jauncey 1975 and references therein). 
This analysis, along with a more detailed study of the number counts 
in the CDF-N, will be presented in G.P. Garmire et~al., in preparation. 

We have also experimented with somewhat less conservative source-selection 
methods than those used above, and these give $N(>S)$ curves that are
consistent with those presented above to within the expected
statistical uncertainties. In the hard band, in fact, varying the
source selection method gives almost identical results. 

In Figure~14 we compare our $N(>S)$ curves with those from 
Mushotzky et~al. (2000) and Tozzi et~al. (2001). In the soft band,
our $N(>S)$ curve is above those of both Mushotzky et~al. (2000) 
and Tozzi et~al. (2001), even at quite bright X-ray fluxes.
However, the results are not discrepant given the error bars, and
any small differences could be due to field-to-field ``cosmic variance.'' 
In the hard band, our $N(>S)$ curve is consistent with those of 
Mushotzky et~al. (2000) and Tozzi et~al. (2001) at fluxes above 
$3\times 10^{-15}$~erg~cm$^{-2}$~s$^{-1}$, and at fainter fluxes
it is intermediate between the two curves. 

\subsection{Extended-Source Detection and Results}

We have searched the standard-band images for extended X-ray sources 
using the CXC's Voronoi Tessellation and Percolation algorithm 
{\sc vtpdetect} (Ebeling \& Wiedenmann 1993; Dobrzycki et~al. 1999). 
We have used a false-positive probability threshold of $1\times 10^{-7}$, 
and we require at least 50 counts per source. Extended source detections 
were checked by inspection of adaptively smoothed images. 

In Figure~15 we show the two most significant extended sources revealed by
our source detection: CXOHDFN~J123620.0+621554 and CXOHDFN~J123756.0+621506. 
Both of these sources are detected most clearly in the soft-band image. 
We have checked for possible problems with these source detections
and find none. The exposure map is relatively smooth near both of 
these sources, so background gradients should not have confused the
extended source detection. In addition, both of these sources are 
visible in adaptively smoothed soft-band images made using only data 
with roll angles of 36.4--44.5$^\circ$ or 134.3--143.8$^\circ$ (see 
Table~1); this argues against an instrumental origin of these sources. 
CXOHDFN~J123620.0+621554 has $274\pm 37$ soft-band counts and no obvious 
optical counterpart in the $I$-band image of Barger et~al. (1999), suggesting 
a fairly high redshift group or cluster. The counts for extended sources
here and hereafter were determined with manual aperture photometry
excluding point sources and regions of strongly varying background; 
for CXOHDFN~J123620.0+621554 we used an elliptical aperture with
semimajor axis $45\arcsec$, semiminor axis $25\arcsec$, and 
position angle $145^\circ$. 
CXOHDFN~J123756.0+621506 has $303\pm 26$ soft-band counts
(in an elliptical aperture with semimajor axis $25\arcsec$, 
semiminor axis $18\arcsec$, and position angle $55^\circ$). 
In the $R$-band image of Liu et~al. (1999), it coincides with a pair 
of optically bright galaxies that appear to be interacting; the X-ray
source is likely to be a low-to-moderate redshift group.

One additional extended source detected by {\sc vtpdetect} with lower
significance than the two above is CXOHDFN~J123645.0+621142
(see Figure~16); we consider this source to be marginally detected. 
This source is notable because it is in the HDF-N itself near the 
$z=1.013$ Fanaroff-Riley~I (FR~I) radio galaxy VLA~J123644.3+621133 
and a number of other $z\approx 1.01$ objects (e.g., Richards et~al. 1998). 
In Paper~IV we detected the FR~I (as CXOHDFN~J123644.3+621132) and 
searched for cluster X-ray emission in its vicinity because FR~I 
sources are often located in clusters of galaxies. No cluster 
emission was found, but with the additional data here it appears 
likely that such emission was present just below the detection 
threshold attainable in Paper~IV. The FR~I lies within the 
emission but appears offset from its (poorly defined) center. 
CXOHDFN~J123645.0+621142 has $\approx 100$ soft-band counts
(in a circular aperture with radius $30\arcsec$). 
Because there are substantially more counts in the putative 
cluster emission than from the FR~I, we are confident that 
this emission is not merely due to counts from the FR~I in the 
wings of the PSF. 

Inspection of the raw and adaptively smoothed soft-band images 
revealed the likely presence of diffuse emission near the
positions 123557.0+621551, 123704.6+621652, and 123721.2+621526. 
Possible groups or clusters are apparent in the 
$R$-band image of Liu et~al. (1999) or the 
$I$-band image of Barger et~al. (1999) near these 
positions. Optical spectroscopy is required for 
confirmation.

Dawson et~al. (2001) have discovered a $z=0.85$ cluster 
centered near 
\hbox{$\alpha_{2000}=$~12$^{\rm h}$ 36$^{\rm m}$ 39\fs 6},
\hbox{$\delta_{2000}=$~$+62^\circ$15$^{\prime}$54$^{\prime\prime}$}. 
Unfortunately, this position is near one of the strongest 
CCD gap features in the exposure map (see Figure~7).
Inspection of the adaptively smoothed soft-band image
for diffuse emission shows a possible $\approx 50$ event 
enhancement near cluster member F~36397+1547 (see Table~4
of Dawson et~al. 2001), but this is not formally detected 
by {\sc vtpdetect} and could be due to a faint point 
source. We do clearly detect X-ray emission from cluster 
member F~36421+1545; this source appears to be pointlike. 

We have also inspected the \chandra\ images near the Wide Angle 
Tail (WAT) radio galaxy VLA~J123725.7+621128 (Muxlow et~al. 1999) since 
WATs are often found in clusters of galaxies. We find no hint of any 
extended X-ray emission centered near this source. 

Extended sources will be discussed in further detail in F.E. Bauer et~al., 
in preparation.


\section{Conclusions and Summary}

These CDF-N observations will contribute greatly to 
the resolution of some of the outstanding questions in extragalactic 
X-ray astronomy. Near the aim point, they 
have reached soft-band and hard-band source densities of
$7100^{+1100}_{-940}$~deg$^{-2}$ (at \hbox{$4.2\times 10^{-17}$~erg~cm$^{-2}$~s$^{-1}$}) and 
$4200^{+670}_{-580}$~deg$^{-2}$ (at \hbox{$3.8\times 10^{-16}$~erg~cm$^{-2}$~s$^{-1}$}), 
respectively. This paper has described the details of the 
observations, data reduction, and technical analysis and presented 
1~Ms source catalogs. It will require many years of work to 
investigate all of the presented sources in detail. This 
survey should remain one of the deepest ever made from 0.5--8.0~keV 
until missions such as the {\it X-ray Evolving Universe Spectroscopy\/} 
(\xeus) mission begin operation,\footnote{For details 
on \xeus\ see http://astro.estec.esa.nl/SA-general/Projects/XEUS/.}
and we hope to extend this survey to an exposure of 5--10~Ms given
the continued operation of \chandra; we have been allocated an 
additional 1~Ms in \chandra\ Cycle~3 (all data will become public
immediately). The current 1~Ms survey is far from the limit
of \chandra's capability. The detector
background is so low that, with appropriate grade screening, 
we will not fully enter the background-limited regime near the
aim point for exposure times of $\simlt~5$~Ms (for the full
band; the situation is even better for the soft band due to 
its substantially lower background). A 5~Ms 
exposure would achieve soft-band and hard-band sensitivities
of $\approx~6\times 10^{-18}$~erg~cm$^{-2}$~s$^{-1}$ and
$\approx~4\times 10^{-17}$~erg~cm$^{-2}$~s$^{-1}$, respectively,
as well as provide detailed spectral, variability, and morphological
constraints on the sources in the present catalog. It would 
provide key information on the existence and nature of the 
sources to be targeted by future missions such as \xeus,
laying the groundwork for these missions.  
\xmm\ will perform deeper surveys in the $\approx$~8--12~keV band, 
but cross-correlation of the sources found by \xmm\ with those 
presented here will still be important to refine the \xmm\
positions and minimize source confusion. 

The source catalogs and images shown in Figure~3
are available on the World Wide Web.\footnote{See
http://www.astro.psu.edu/users/niel/hdf/hdf-chandra.html.}
We will continue to improve the source catalogs as better 
calibration information, analysis methods and software become 
available. For example, we plan to optimize the searching
for variable sources and search more sensitively for X-ray 
sources that correlate with sources at other wavelengths. 
Improved searching for diffuse sources will be performed as
the \chandra\ background becomes better understood. 


\acknowledgments

This work would not have been possible without the enormous efforts 
of the entire \chandra\ and ACIS teams. 
We thank P.E.~Freeman, J.~Gaffney, D.E.~Harris, M.~Karovska, C.~Liu,
and P.~Zhao for helpful discussions and sharing data. 
We gratefully acknowledge the financial support of
NASA grant NAS~8-38252 (GPG, PI),
NSF CAREER award AST-9983783 (WNB, DMA, FEB),  
NASA GSRP grant NGT~5-50247 and the Pennsylvania Space Grant Consortium (AEH), 
NSF grant AST-9900703~(DPS),  
NASA Hubble Fellowship grant HF-01117.01-A and NSF grant AST-0084847 (AJB), and 
NSF grant AST-0084816 (LLC).


\clearpage


\begin{deluxetable}{lccclllcl}
\tabletypesize{\tiny}
\tablewidth{0pt}
\tablecaption {Journal of \chandra\ Observations}
%
\scriptsize
\tablehead{
\colhead{Obs.}                                 &
\colhead{Obs.}                                 &
\colhead{Exposure}                             &
\multicolumn{2}{c}{Aim Point$^{\rm b}$}        &
\multicolumn{2}{c}{ACIS-I Corners}             &
\colhead{Roll}                                 & 
\colhead{Pipeline}                             \\
\colhead{ID}                                 &
\colhead{Start}                              &
\colhead{Time (s)$^{\rm a}$ }                &
\colhead{$\alpha_{\rm 2000}$}                &
\colhead{$\delta_{\rm 2000}$}                &
\colhead{$\alpha_{\rm 2000}$}                &
\colhead{$\delta_{\rm 2000}$}                &
\colhead{Angle ($^\circ$)$^{\rm c}$}         &
\colhead{Version$^{\rm d}$}         
}
\startdata
580  & 1999 Nov 13 &  49437 & 12 37 13.1 & +62 12 42   & 12 37 35.0 & +62 24 16 & 36.4   & R4CU5UPD9 \\
     & 01:14       &        &            &             & 12 35 39.8 & +62 14 10 &  & \\
     &             &        &            &             & 12 37 05.9 & +62 00 37 &  & \\
     &             &        &            &             & 12 39 02.6 & +62 10 32 &  & \\
\\   
967  & 1999 Nov 14 &  57440 & 12 37 01.0 & +62 12 57   & 12 37 20.6 & +62 24 36 & 38.0   & R4CU5UPD9 \\
     & 19:47       &        &            &             & 12 35 27.9 & +62 14 16 &  & \\
     &             &        &            &             & 12 36 57.6 & +62 00 53 &  & \\
     &             &        &            &             & 12 38 51.4 & +62 11 07 &  & \\
\\  
966  & 1999 Nov 21 &  57443 & 12 37 01.1 & +62 12 57   & 12 37 19.7 & +62 24 33 & 38.9   & R4CU5UPD9 \\
     & 04:03       &        &            &             & 12 35 27.6 & +62 13 55 &  & \\
     &             &        &            &             & 12 36 58.8 & +62 00 54 &  & \\
     &             &        &            &             & 12 38 51.6 & +62 11 19 &  & \\
\\  
957  & 2000 Feb 23 &  57446 & 12 36 36.4 & +62 14 44   & 12 36 51.9 & +62 28 35 & 134.3  & R4CU5UPD12.1 \\
     & 06:31       &        &            &             & 12 35 06.2 & +62 16 43 &  & \\
     &             &        &            &             & 12 36 47.8 & +62 04 36 &  & \\
     &             &        &            &             & 12 38 33.4 & +62 16 27 &  & \\
\\
2386 & 2000 Nov 20 &   9839 & 12 37 04.3 & +62 13 04   & 12 37 09.7 & +62 24 55 & 44.5   & R4CU5UPD11.2 \\
     & 05:39       &        &            &             & 12 35 26.7 & +62 12 58 &  & \\
     &             &        &            &             & 12 37 09.0 & +62 01 04 &  & \\
     &             &        &            &             & 12 38 52.7 & +62 12 42 &  & \\
\\
1671 & 2000 Nov 21 & 166833 & 12 37 04.3 & +62 13 04   & 12 37 09.4 & +62 24 52 & 44.5   & R4CU5UPD13.1 \\
     & 13:26       &        &            &             & 12 35 25.7 & +62 12 58 &  & \\
     &             &        &            &             & 12 37 08.2 & +62 00 53 &  & \\
     &             &        &            &             & 12 38 51.8 & +62 12 44 &  & \\
\\
2344 & 2000 Nov 24 &  90836 & 12 37 04.3 & +62 13 04   & 12 37 09.4 & +62 24 56 & 44.5   & R4CU5UPD12.1 \\
     & 05:41       &        &            &             & 12 35 25.5 & +62 12 57 &  & \\
     &             &        &            &             & 12 37 08.1 & +62 00 49 &  & \\
     &             &        &            &             & 12 38 52.8 & +62 12 40 &  & \\
\\
2232 & 2001 Feb 19 & 129223 & 12 36 35.8 & +62 14 40   & 12 36 46.2 & +62 28 44 & 136.8  & R4CU5UPD14.1 \\
     & 14:24       &        &            &             & 12 35 05.6 & +62 16 15 &  & \\
     &             &        &            &             & 12 36 51.7 & +62 04 35 &  & \\
     &             &        &            &             & 12 38 31.9 & +62 17 05 &  & \\
\\
2233 & 2001 Feb 22 &  60813 & 12 36 35.4 & +62 14 37   & 12 36 42.8 & +62 28 38 & 138.3  & R4CU5UPD14.1 \\
     & 03:44       &        &            &             & 12 35 04.9 & +62 15 53 &  & \\
     &             &        &            &             & 12 36 52.9 & +62 04 41 &  & \\
     &             &        &            &             & 12 38 30.8 & +62 17 20 &  & \\
\\
2423 & 2001 Feb 23 &  68420 & 12 36 35.4 & +62 14 37   & 12 36 42.7 & +62 28 45 & 138.3  & R4CU5UPD14.1 \\
     & 06:57       &        &            &             & 12 35 05.5 & +62 15 58 &  & \\
     &             &        &            &             & 12 36 54.0 & +62 04 42 &  & \\
     &             &        &            &             & 12 38 31.5 & +62 17 25 &  & \\
\\
2234 & 2001 Mar 02 & 165924 & 12 36 34.5 & +62 14 30   & 12 36 33.1 & +62 28 41 & 142.3  & R4CU5UPD14.1 \\
     & 2:00        &        &            &             & 12 35 02.9 & +62 15 06 &  & \\
     &             &        &            &             & 12 36 58.7 & +62 04 42 &  & \\
     &             &        &            &             & 12 38 29.4 & +62 18 10 &  & \\
\\
2421 & 2001 Mar 04 &  61647 & 12 36 34.1 & +62 14 27   & 12 36 28.9 & +62 28 38 & 143.8  & R4CU5UPD14.1 \\
     & 16:52       &        &            &             & 12 35 02.8 & +62 14 46 &  & \\
     &             &        &            &             & 12 36 59.8 & +62 04 50 &  & \\
     &             &        &            &             & 12 38 27.4 & +62 18 23 &  & \\
\enddata
\tablecomments{The focal-plane temperature was 
$-110^\circ$\,C during the first three observations and 
$-120^\circ$\,C during the others.}
\tablenotetext{a}{All observations were continuous. These times have been corrected for 
lost exposure due to the 0.041~s read-out time per CCD frame, and we have also removed
short time intervals with bad satellite aspect.}
\tablenotetext{b}{The average aim point, weighted by exposure time, is
$\alpha_{2000}=$~12$^{\rm h}$ 36$^{\rm m}$ $48\fs 1$,
$\delta_{2000}=$~$+62^\circ$13$^{\prime}$53$^{\prime\prime}$.} 
\tablenotetext{c}{Roll angle describes the orientation of the \chandra\ instruments
on the sky.  The angle is between 0--360$^{\circ}$, and it increases to the West
of North (opposite to the sense of traditional position angle).}
\tablenotetext{d}{This is the version of the CXC pipeline software used for basic 
processing of the data.}
\end{deluxetable}

\clearpage


\begin{deluxetable}{lll}
\tabletypesize{\small}
\tablewidth{0pt}
\tablecaption {Grade Sets}
\tablehead{
\colhead{Name}                            &                                
\colhead{Band}                            &                                
\colhead{Grades}                          
}
\startdata
Standard \asca\   & Full (0.5--8.0~keV)   & \asca\ grades 0, 2, 3, 4, 6   \\
grade set         & Soft (0.5--2.0~keV)   & \asca\ grades 0, 2, 3, 4, 6   \\
                  & Hard (2--8~keV)       & \asca\ grades 0, 2, 3, 4, 6   \\
                  & Ultrahard (4--8~keV)  & \asca\ grades 0, 2, 3, 4, 6   \\
                  &                       &                               \\
Restricted ACIS   & Full (0.5--8.0~keV)   & ACIS grades 0, 2, 8, 16, 64   \\
grade set         & Soft (0.5--2.0~keV)   & ACIS grades 0, 64             \\
                  & Hard (2--8~keV)       & ACIS grades 0, 2, 8, 16       \\
                  & Ultrahard (4--8~keV)  & ACIS grades 0, 2, 8, 16       \\
\enddata
\end{deluxetable}


\begin{deluxetable}{lll}
\tablecaption {For this table please see the World Wide Web site listed in Footnote~13.}
\tablehead{
\colhead{ } &                                
\colhead{ } &                                
\colhead{ }                          
}
\startdata
  & & \\
\enddata
\end{deluxetable}

\clearpage


\setcounter{table}{3}

\begin{deluxetable}{lccccc}
\tabletypesize{\small}
\tablewidth{0pt}
\tablecaption {Summary of \chandra\ Source Detections}
\scriptsize
\tablehead{
\colhead{Energy}                                 &
\colhead{Number of}                              &
\multicolumn{4}{c}{Detected Counts Per Source}   \\
\colhead{Band}                                      &
\colhead{Sources$^{\rm a}$ }                        &
\colhead{Maximum}                                   &
\colhead{Minimum}                                   &
\colhead{Median}                                    &
\colhead{Mean}                                      
}
\startdata
Full          & 360 & 9100.7 & 7.8 & 75.8 & 289.0  \\
Soft          & 325 & 6633.0 & 5.5 & 44.6 & 211.5  \\
Hard          & 265 & 2481.0 & 6.0 & 50.7 & 130.2  \\
Ultrahard     & 145 & 792.0  & 6.4 & 36.0 & 69.2   \\
\enddata
\tablenotetext{a}{There are 370 independent X-ray sources detected in total
with a false-positive probability threshold of $1\times 10^{-7}$. We have 
included cross-band counterparts from the {\sc wavdetect} runs with a 
false-positive probability threshold of $1\times 10^{-5}$ (see \S3.2.1).}
\end{deluxetable}


\begin{deluxetable}{lcccc}
\tabletypesize{\small}
\tablewidth{0pt}
\tablecaption {Sources Detected in One Band but Not Another}
\scriptsize
\tablehead{
\colhead{Detection}                              &
\multicolumn{4}{c}{Non-Detection Energy Band}    \\
\colhead{Energy Band}                            &
\colhead{Full}                                   &
\colhead{Soft}                                   &
\colhead{Hard}                                   &
\colhead{Ultrahard}                                      
}
\startdata
Full          & 0  & 45 & 95 & 215   \\
Soft          & 10 & 0  & 98 & 202   \\
Hard          & 0  & 38 & 0  & 121   \\
Ultrahard     & 0  & 22 & 1  & 0     \\
\enddata
\tablecomments{For example, there were 45 sources detected in the
full band but not in the soft band.}
%
\end{deluxetable}


\begin{deluxetable}{lll}
\tablecaption {For this table please see the World Wide Web site listed in Footnote~13.}
\tablehead{
\colhead{ } &                                
\colhead{ } &                                
\colhead{ }                          
}
\startdata
  & & \\
\enddata
\end{deluxetable}



\setcounter{table}{6}

\begin{deluxetable}{lccc}
\tabletypesize{\small}
\tablewidth{0pt}
\tablecaption {Source Selection for the Number Counts}
\tablehead{
\colhead{Flux Range}                      &                                
\colhead{Maximum Source}                  &                                
\colhead{Minimum Effective}               & 
\colhead{Selected Solid}                  \\
\colhead{(erg~cm$^{-2}$~s$^{-1}$)}        &                                
\colhead{Selection Radius$^{\rm a}$}      &                                
\colhead{Exposure Time (ks)}              & 
\colhead{Angle (arcmin$^2$)$^{\rm b}$}                          
}
\startdata
\hline
\multicolumn{4}{c}{Soft Band} \\ \hline
(4.2--20)$\times 10^{-17}$ & $3^{\prime}$  & $870$ & 18.75  \\
$>2\times 10^{-16}$        & $5^{\prime}$  & $870$ & 50.26  \\
\hline
\multicolumn{4}{c}{Hard Band} \\ \hline
(3.8--8)$\times 10^{-16}$ & $3^{\prime}$  & $870$ & 17.89  \\
(8--45)$\times 10^{-16}$  & $5^{\prime}$  & $870$ & 45.58  \\
$>4.5\times 10^{-15}$     & $7^{\prime}$  & $840$ & 91.64  \\
\enddata
\tablenotetext{a}{Only sources with angular separations from the average
aim point smaller than this value were used for the number counts 
calculations in the corresponding flux range.}
\tablenotetext{b}{Amount of solid angle satisfying the spatial 
and effective exposure time filtering criteria.}
\end{deluxetable}

\clearpage


\begin{figure}
\vspace{-0.5truein}
\epsscale{0.8}
\figurenum{1}
\plotone{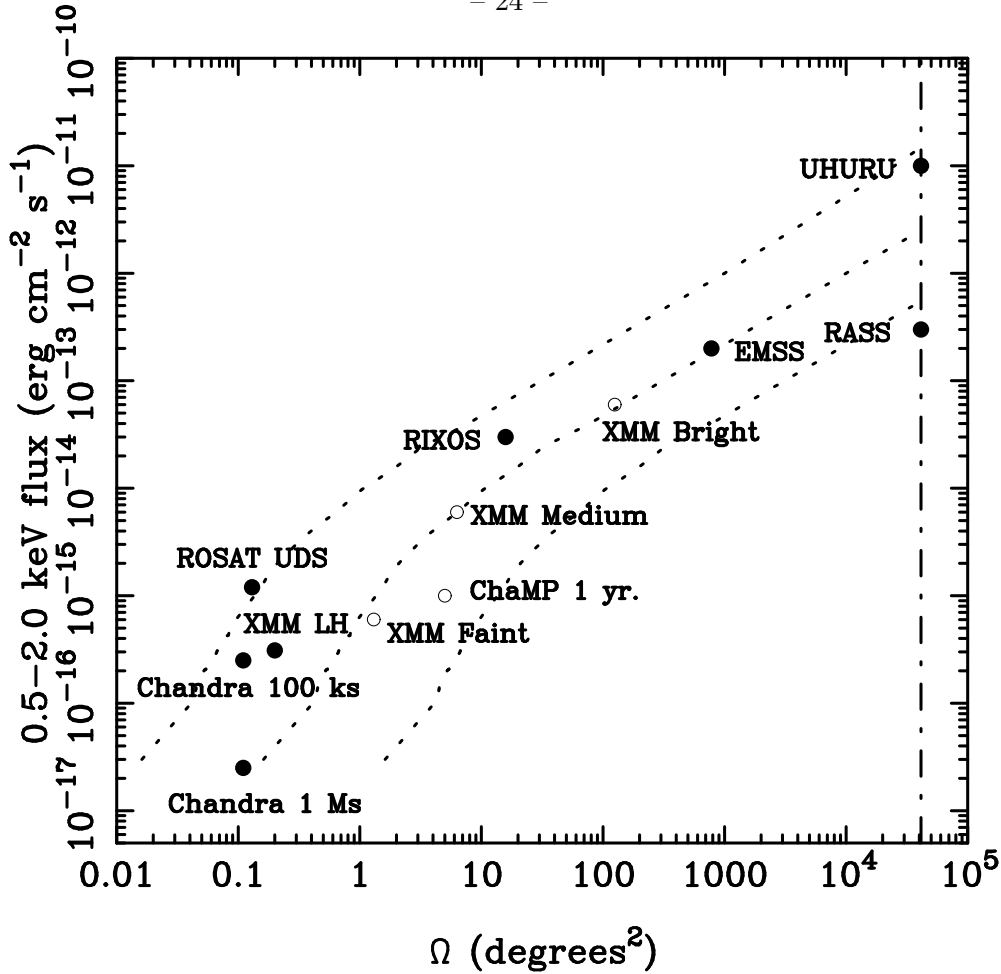}
\vspace{-0.1truein}
\caption{Distribution of some extragalactic X-ray surveys in the 0.5--2.0~keV 
flux limit versus solid angle, $\Omega$, plane. Shown are 
the \uhuru\ survey (e.g., Forman et~al. 1978), 
the \rosat\ All-Sky Survey (RASS; e.g., Voges et~al. 1999), 
the \einstein\ Extended Medium-Sensitivity Survey (EMSS; e.g., Gioia et~al. 1990), 
the \rosat\ International X-ray/Optical Survey (RIXOS; e.g., Mason et~al. 2000), 
the \xmm\ Serendipitious Surveys (\xmm\ Bright, \xmm\ Medium, \xmm\ Faint; e.g., Watson et~al. 2001), 
the \chandra\ Multiwavelength Project (ChaMP; e.g., Wilkes et~al. 2001),
the \rosat\ Ultra Deep Survey (UDS; e.g., Lehmann et~al. 2001), 
the deep \xmm\ survey of the Lockman Hole (\xmm\ LH; e.g., Hasinger et~al. 2001), 
\chandra\ 100~ks surveys (e.g., Mushotzky et~al. 2000), and 
\chandra\ 1~Ms surveys (i.e., the \hbox{CDF-N} survey and the \chandra\ 
Deep Field South survey). 
Solid dots are for surveys that have been completed, and open circles
are for surveys that are in progress. 
Clearly, each of the surveys shown has a range of flux limits across its
solid angle (due to effects such as off-axis PSF broadening and vignetting); we 
have generally shown the most sensitive flux limit. 
The dotted curves show, from top to bottom, the loci of 100, 1000, and 
10000 0.5--2.0~keV sources (these have been calculated using the
number counts of Hasinger et~al. 1998, Paper~III, and \S3.2.4); for 
example, a 1~degree$^2$ survey with a 0.5--2.0~keV flux limit of 
$9.5\times 10^{-15}$~erg~cm$^{-2}$~s$^{-1}$ will detect
$\approx 100$ sources. 
The vertical dot-dashed line shows the solid angle of the whole sky.}
\end{figure}


\begin{figure}
\epsscale{0.7}
\figurenum{2}
\plotone{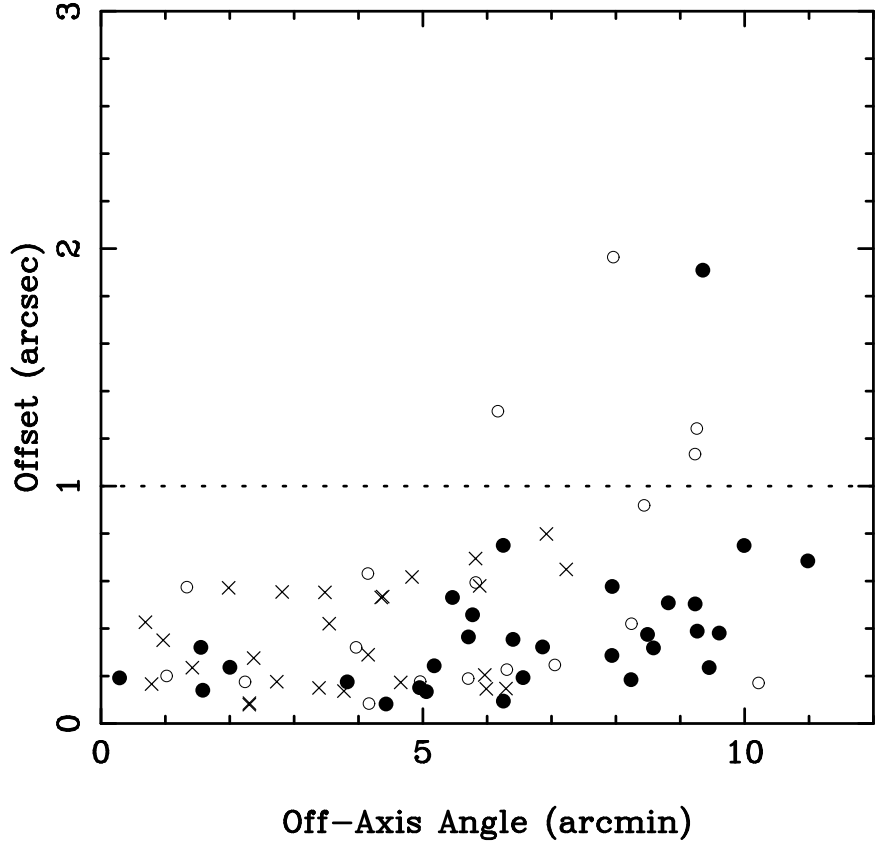}
\vspace{-0.1truein}
\caption{Positional offset versus off-axis angle for full-band \chandra\ 
sources that match with 1.4~GHz sources (from Richards 2000) to within 
$2\farcs 5$. The crosses are \chandra\ sources with 10--50 counts, the 
open circles are \chandra\ sources with 50--100 counts, and the solid 
dots are \chandra\ sources with $>100$ counts.
The X-ray positions are usually good to within $\approx 0\farcs 6$ for 
off-axis angles $<5^{\prime}$. At larger off-axis angles, where the 
HRMA PSF rapidly broadens and becomes complex, 
the positional accuracy, as expected, degrades.
The \chandra\ source positions have been determined following \S3.2.1; 
positions from no-PSF runs were used whenever possible.}
\end{figure}


\begin{figure}
\vspace{-0.5truein}
\epsscale{1.05}
\figurenum{3}
\plotone{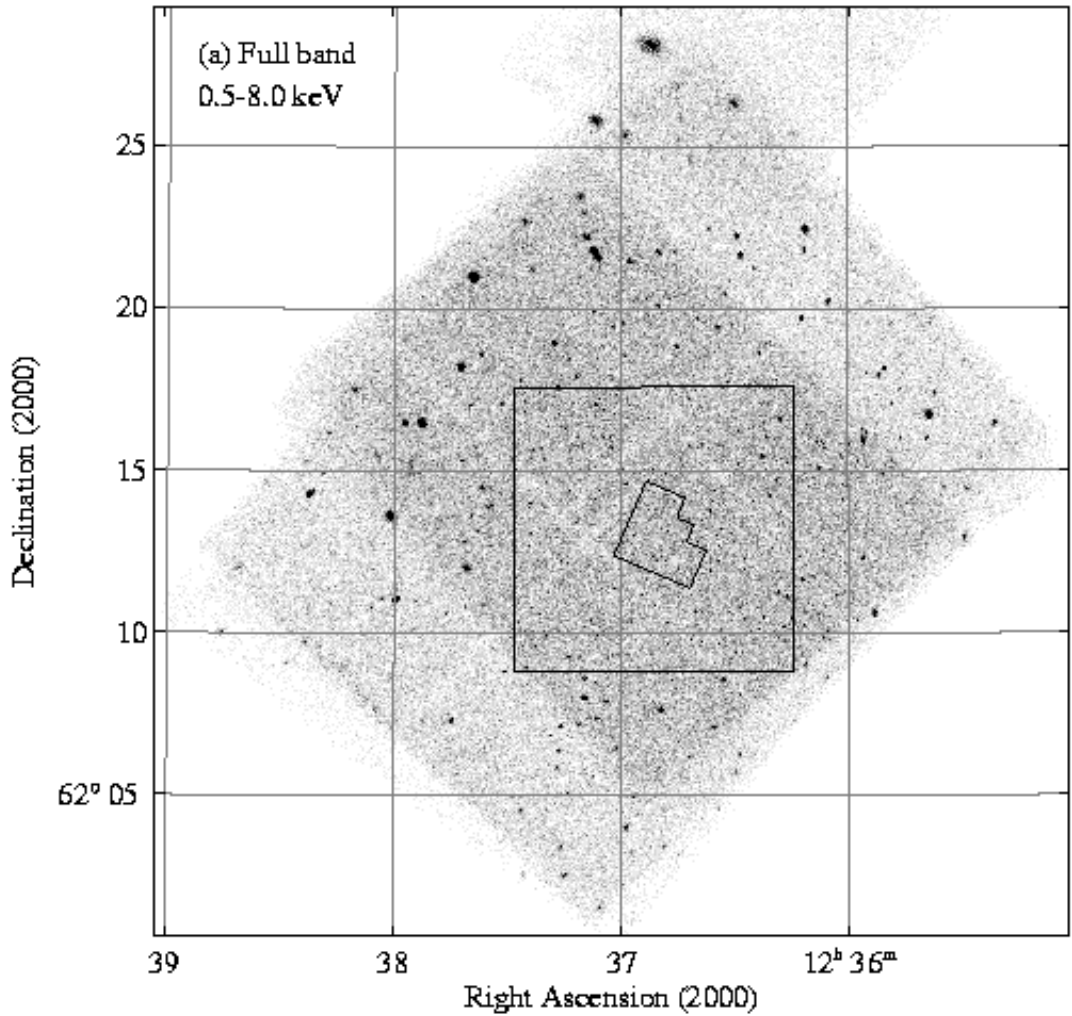}
\vspace{-2.0truein}
\caption{Images of the CDF-N 
in the (a) full band, (b) soft band, (c) hard band, and (d) ultrahard band.
These images have been made using the standard \asca\ grade set (see Table~2),
and they are binned by a factor of four in both right ascension and declination. 
The light grooves running through the images correspond to the gaps 
between the CCDs. The small polygon indicates the HDF-N, and the large 
square indicates the area covered by the Caltech Faint Field Galaxy Redshift 
Survey (e.g., Hogg et~al. 2000; hereafter the ``Caltech area'').
Note: Only one of the four pages of images could be included here; please see
the World Wide Web site listed in Footnote~13 for the version with all
the images.}
\end{figure}





\begin{figure}
\vspace{-0.5truein}
\epsscale{1.05}
\figurenum{4}
\plotone{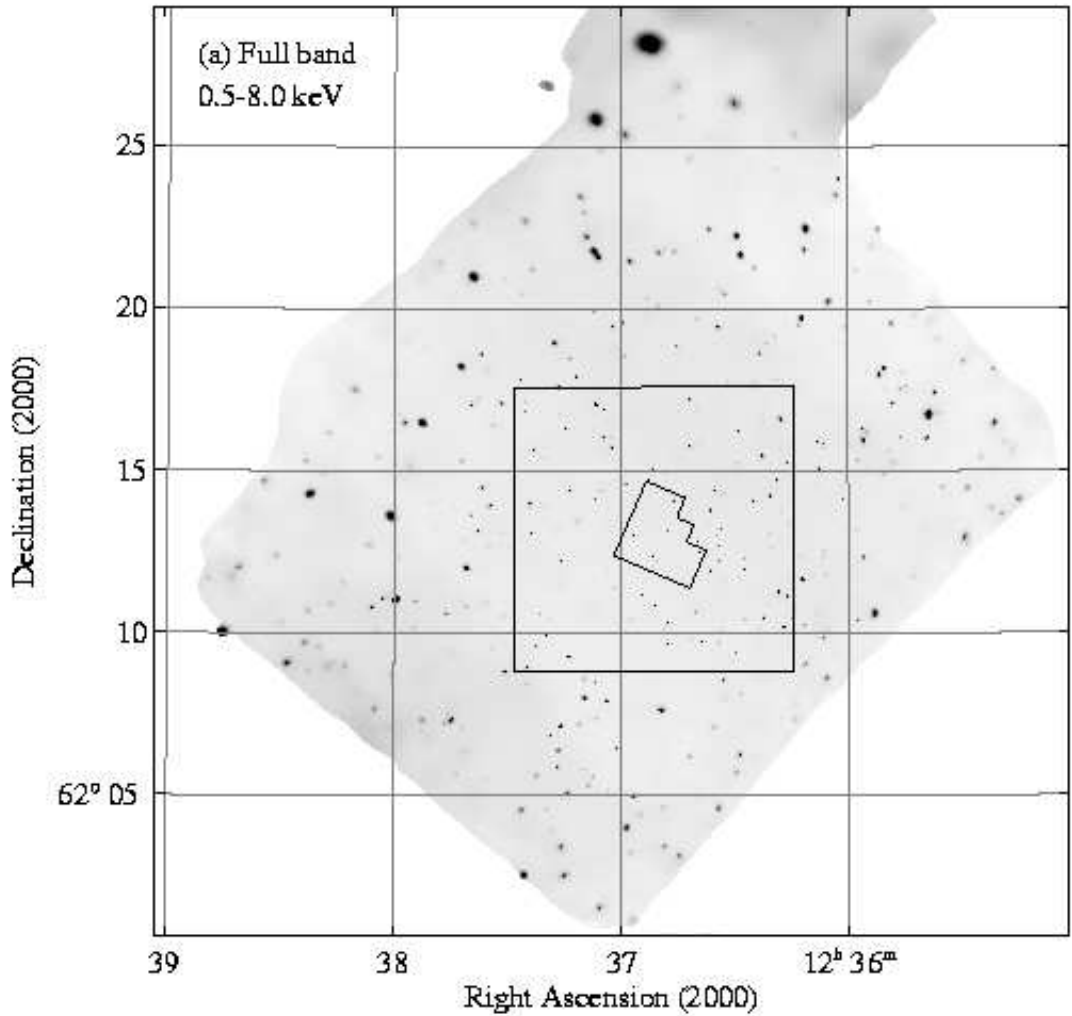}
\vspace{-2.0truein}
\caption{Adaptively smoothed and exposure-map corrected images of the CDF-N 
in the (a) full band, (b) soft band, (c) hard band, and (d) ultrahard band.
These images have been made using the standard \asca\ grade set (see Table~2),
and they are binned by a factor of four in both right ascension and declination. 
The adaptive smoothing has been performed using the code of 
Ebeling, White, \& Rangarajan (2001) at the $2.5\sigma$ level, 
and the grayscales are linear. Much of the apparent diffuse emission
is just instrumental background; see \S3.3 for a discussion of 
extended sources. The edges of the image appear ``rounded'' due 
to the combination of the dither of \chandra\ and the requirement
that the effective exposure time exceed $50$~ks (see \S3.1). 
The small polygon indicates the HDF-N, and the large square indicates 
the Caltech area.
Note: Only one of the four pages of images could be included here; please see
the World Wide Web site listed in Footnote~13 for the version with all
the images.}
\end{figure}





\begin{figure}
\epsscale{1.0}
\figurenum{5}
\plotone{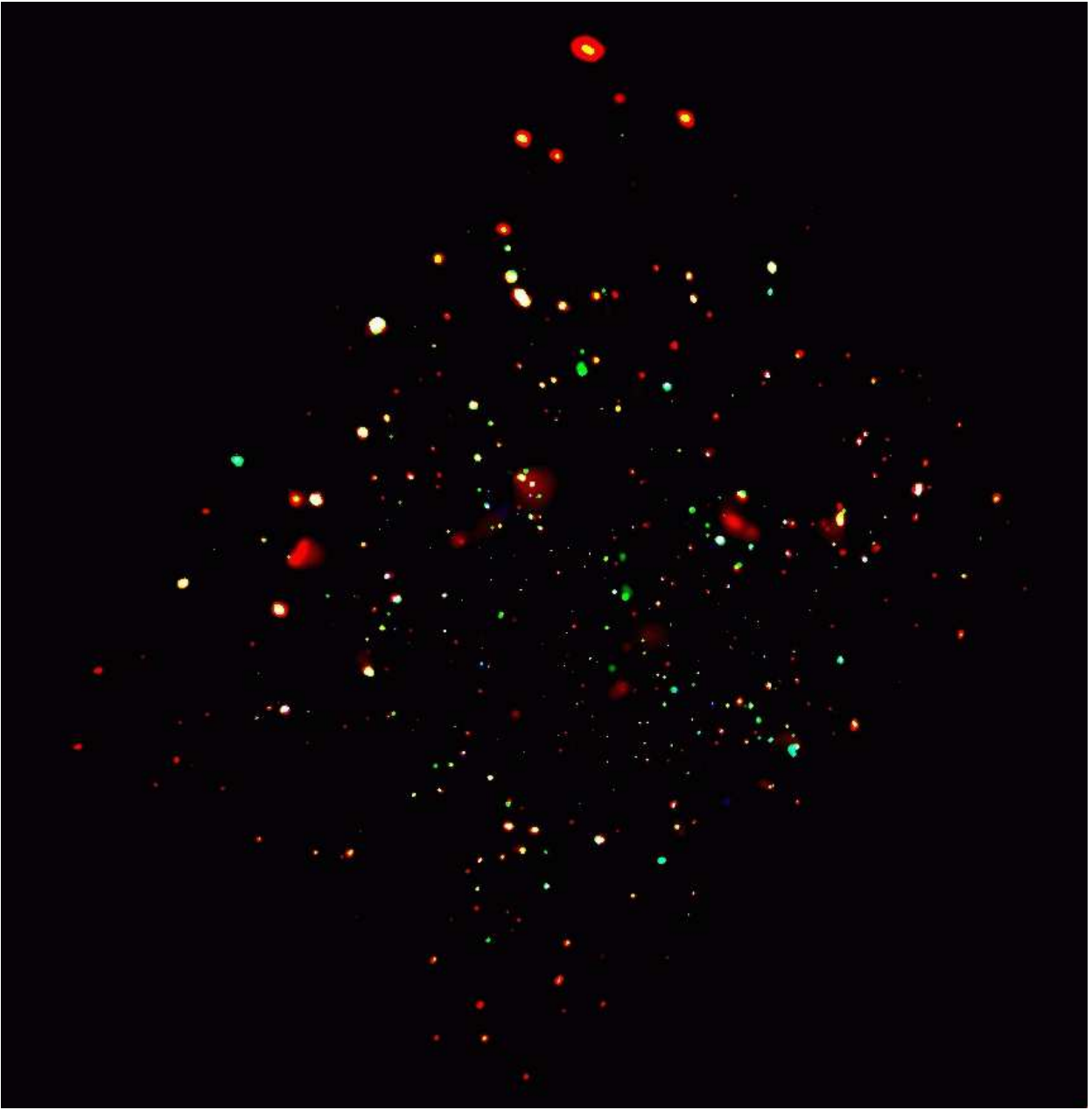}
\vspace{-0.1truein}
\caption{\chandra\ ``true-color'' image of the CDF-N. 
This image has been constructed from the soft-band (red), hard-band (green),
and ultrahard-band (blue) images shown in Figure~4. Two of the red diffuse
patches are CXOHDFN~J123620.0+621554 and CXOHDFN~J123756.0+621506 (see \S3.3).}
\end{figure}


\begin{figure}
\vspace{-1.0truein}
\epsscale{0.9}
\figurenum{6}
\plotone{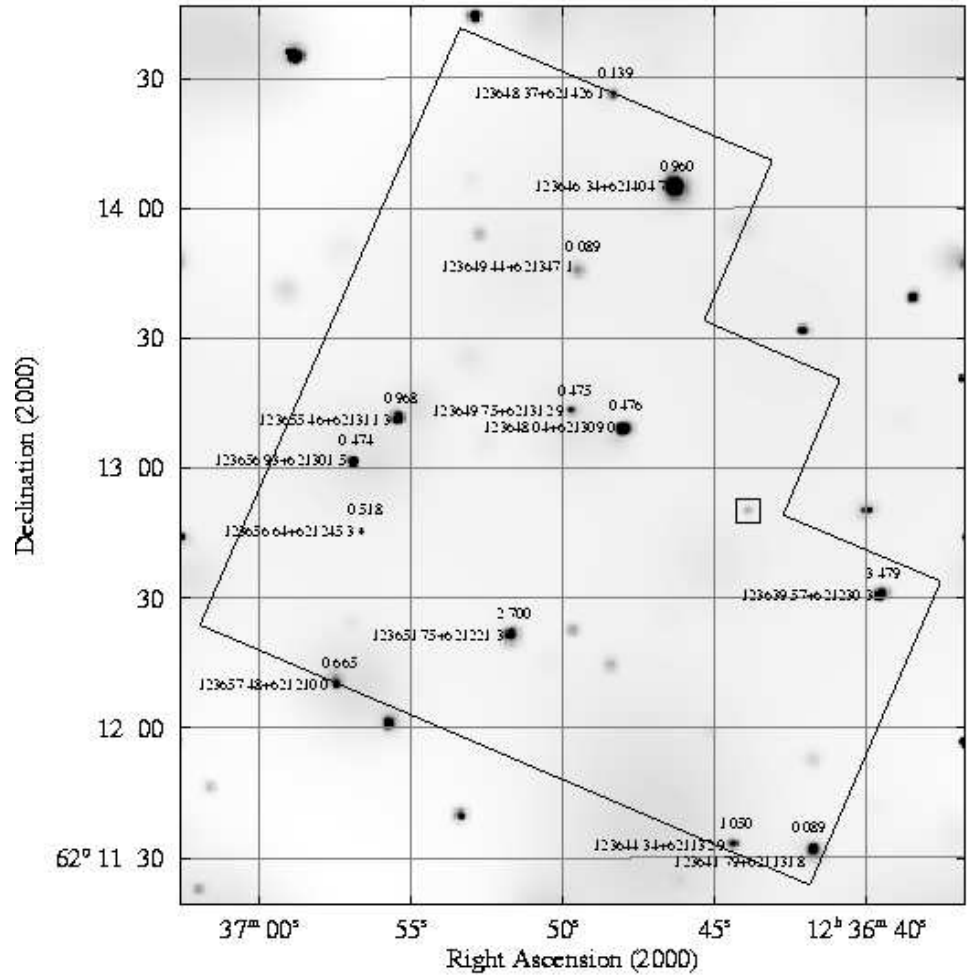}
\vspace{-1.2truein}
\caption{Adaptively smoothed \chandra\ image of the HDF-N and its immediate 
environs in the full band. This image has been made using the standard \asca\ 
grade set, and it has not been binned. The adaptive smoothing has been performed 
using the code of Ebeling et~al. (2001) at the $2.5\sigma$ level, 
and the grayscales are linear. 
Sources detected in this paper are labeled; labels are given immediately 
to the left of the corresponding X-ray sources, and redshifts 
are given immediately above the corresponding X-ray sources. 
The unlabeled enhancements apparent in the HDF-N are discussed
in \S3.2.3. One of these, CXOHDFN~J123643.9+621249, is boxed (this 
is the variable source detected in Paper~IV that is not detected here).}
\end{figure}


\begin{figure}
\vspace{-0.5truein}
\epsscale{1.05}
\figurenum{7}
\plotone{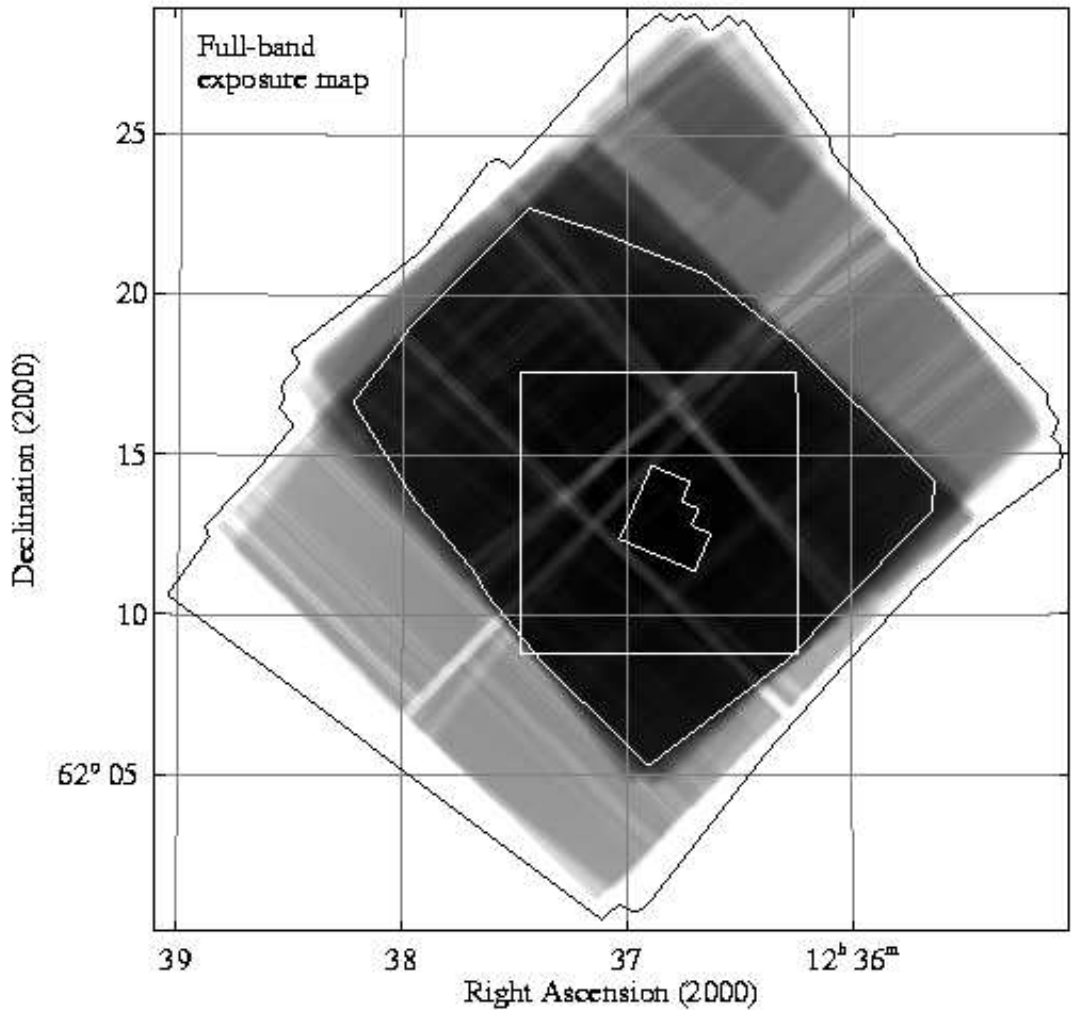}
\vspace{-2.0truein}
\caption{Combined full-band exposure map for the CDF-N ACIS observations. The 
darkest areas correspond to the highest effective exposure times (the 
maximum value is 948.7~ks), and the grayscales are 
logarithmic. The light grooves running through the exposure map 
correspond to the gaps between the CCDs and bad columns. The small white 
polygon indicates the HDF-N itself, the large white square indicates the Caltech area, and 
the large white polygon indicates a ``high-exposure area'' where the median
exposure time is $>800$~ks. The black outline surrounding the exposure map
indicates the extent of all the ACIS-I observations; the regions of the 
exposure map near the outline appear white due to low exposure (20--200~ks).}
\end{figure}


\begin{figure}
\epsscale{0.9}
\figurenum{8}
\plotone{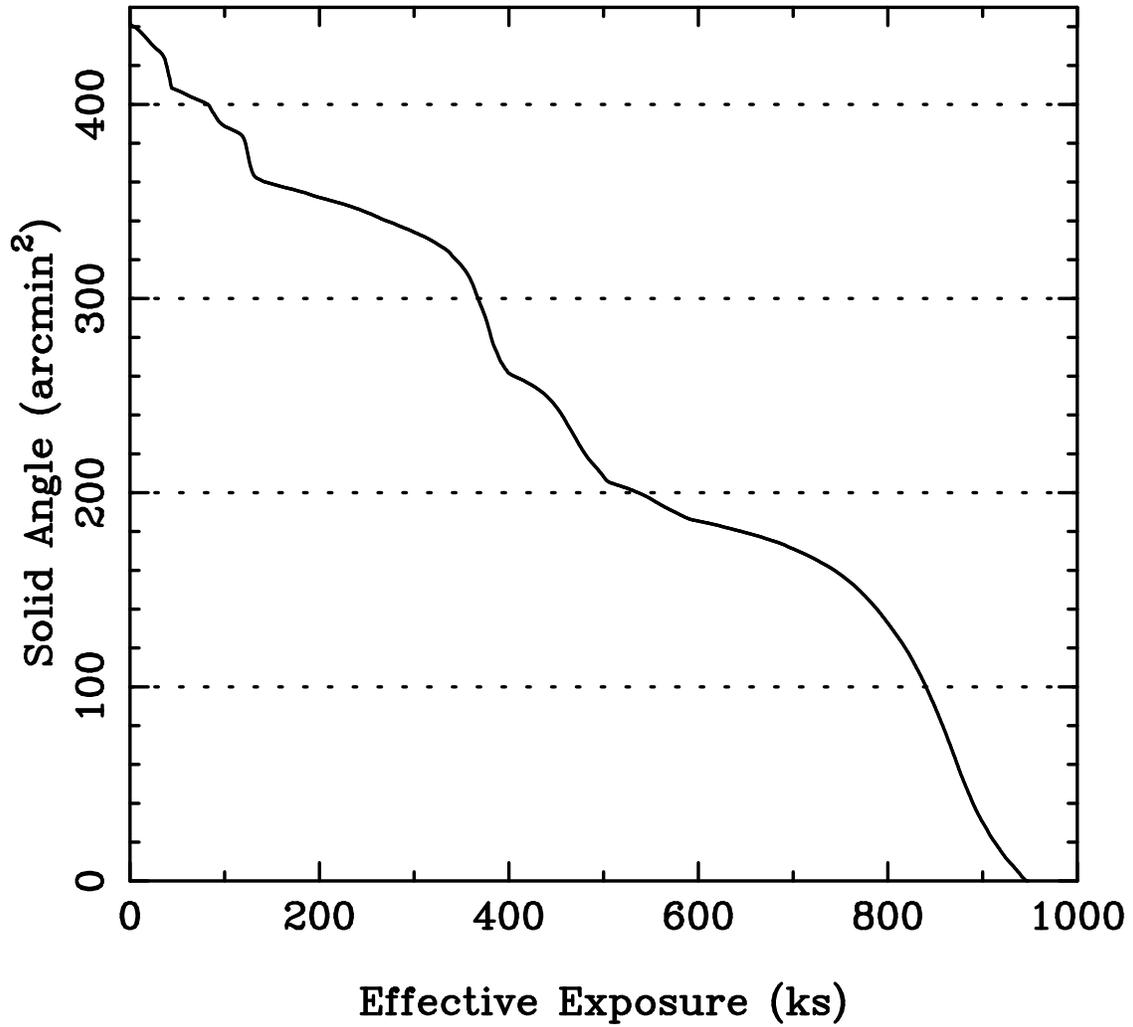}
\vspace{-0.1truein}
\caption{Amount of survey solid angle having at least a given amount of 
effective exposure in the full-band exposure map. Compare with Figure~7.}
\end{figure}

\clearpage


\begin{figure}
\epsscale{0.9}
\figurenum{9}
\plotone{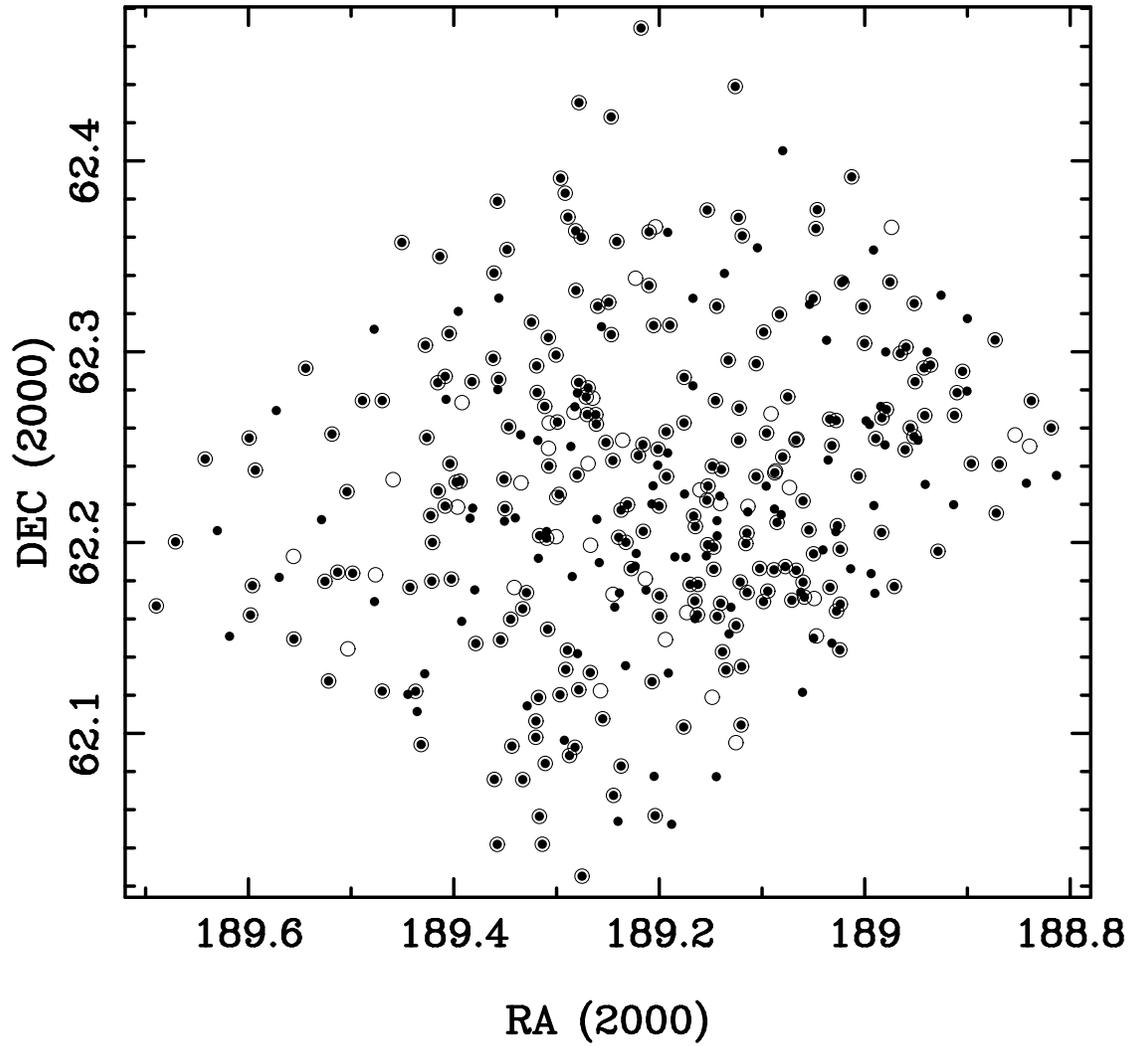}
\vspace{-0.1truein}
\caption{Positions of the soft-band (solid dots) and hard-band (circles) 
sources in Table~3. This plot removes the illusory effect of the 
changing PSF size across the field of view (although, of course, the 
detected source density is still modulated by the varying PSF size
and effective exposure time).}
\end{figure}

\clearpage


\begin{figure}
\epsscale{0.9}
\figurenum{10}
\plotone{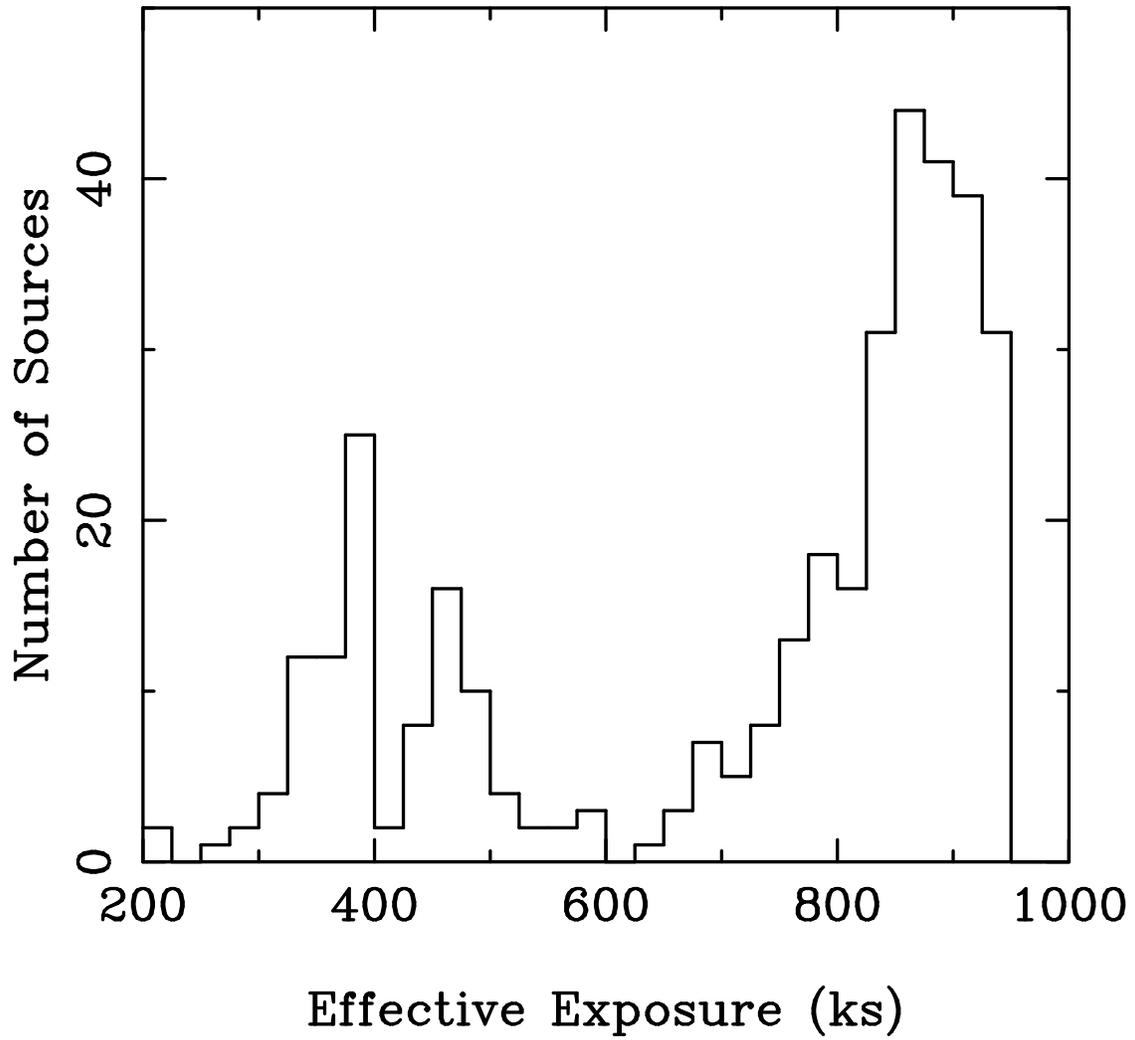}
\vspace{-0.1truein}
\caption{Histogram showing the distribution of full-band effective exposure time 
for the sources in Table~3. Most of the sources have $>600$~ks exposures, but
a significant number have 200--600~ks exposures.}
\end{figure}

\clearpage


\begin{figure}
\epsscale{0.8}
\figurenum{11}
\plotone{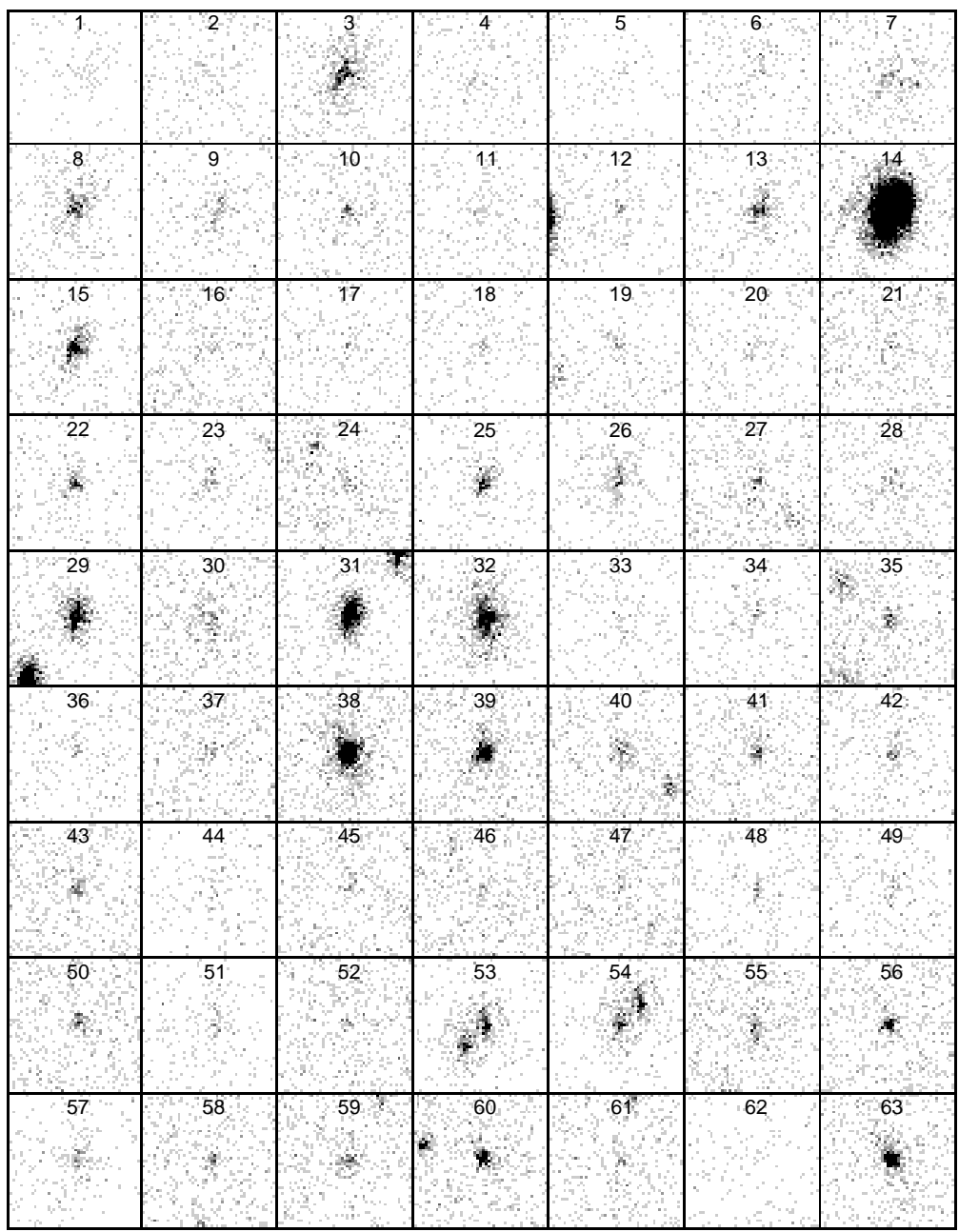}
\vspace{-0.1truein}
\caption{Full-band ``postage-stamp'' images for the \chandra\ sources 
in Table~3; the sources are numbered as in Table~3. Each image is 
oriented so that North is up and East is to the left, and each is
50 pixels ($\approx 24\farcs 6$) on a side. The source of interest
is always at the center of the image. The background varies
significantly from image to image due to the varying effective
exposure time (see Figure~7). 
In a few cases no source is apparent; these sources were not detected in the
full band but were detected in one of the other bands. 
A few of the sources appear to show extent; in most cases this is an artifact 
of the complex \chandra\ off-axis PSF.
Note: Only one of the six pages of cutouts could be included here; please see
the World Wide Web site listed in Footnote~13 for the version with all
the cutouts.}
\end{figure}






\clearpage


\begin{figure}
\epsscale{0.70}
\figurenum{12}
\plotone{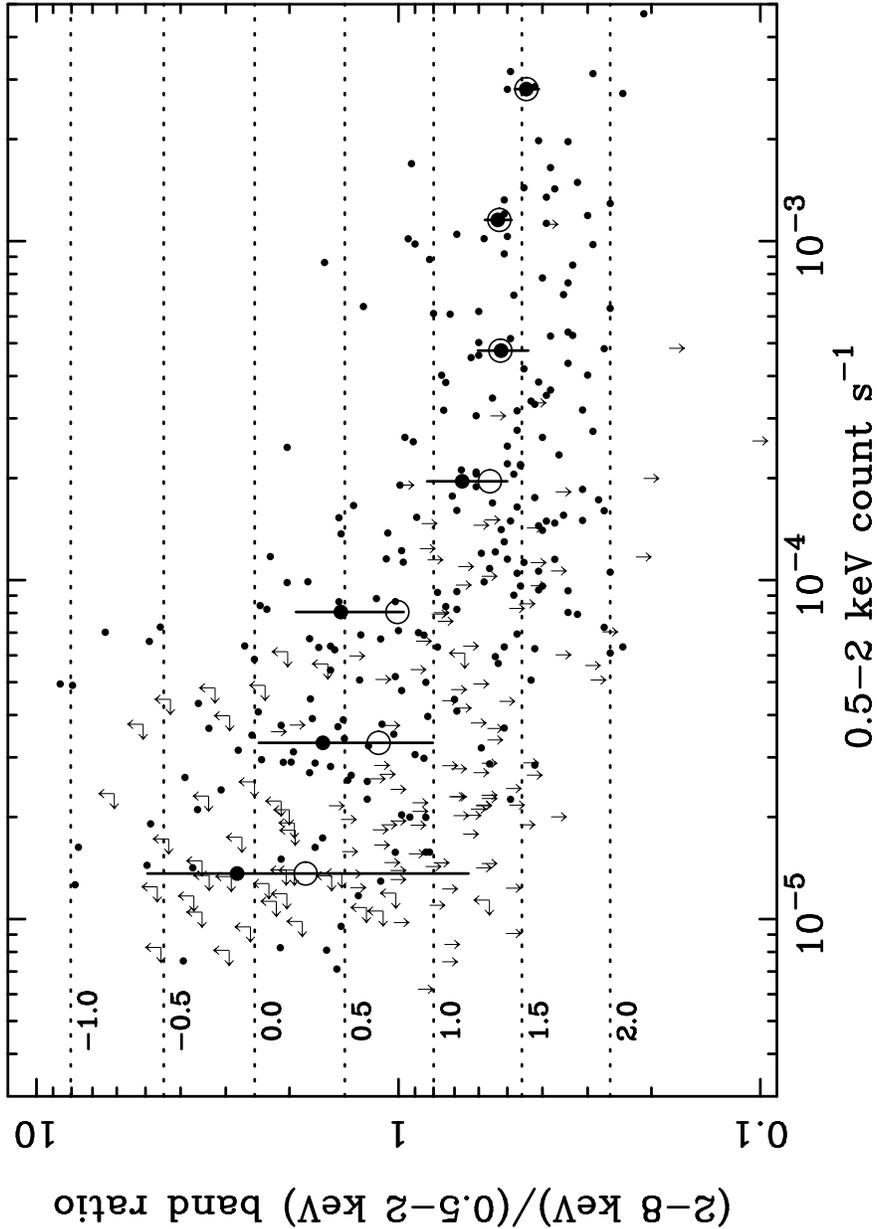}
\vspace{-0.1truein}
\caption{Band ratio as a function of soft-band count rate for the sources
in Table~3.  
Small solid dots show sources detected in both the soft and hard bands, and 
plain arrows show sources detected in only one of these two bands (sources
detected in only the full band are not plotted). To reduce symbol crowding, 
we do not show error bars for each of the small solid dots. Instead, the 
large solid dots show average band ratios and error bars for the small solid 
dots as a function of soft-band count rate (these large solid dots are given
only to show the size of the errors, and they should not be interpreted
statistically). 
The open circles show average band ratios derived from stacking analyses
(following \S3.3 of Paper~VI). 
Horizontal dotted lines are labeled with the photon indices that correspond to 
a given band ratio assuming only Galactic absorption (these were determined 
using the CXC's Portable, Interactive, Multi-Mission Simulator).} 
\end{figure}


\begin{figure}
\epsscale{1.0}
\figurenum{13}
\plotone{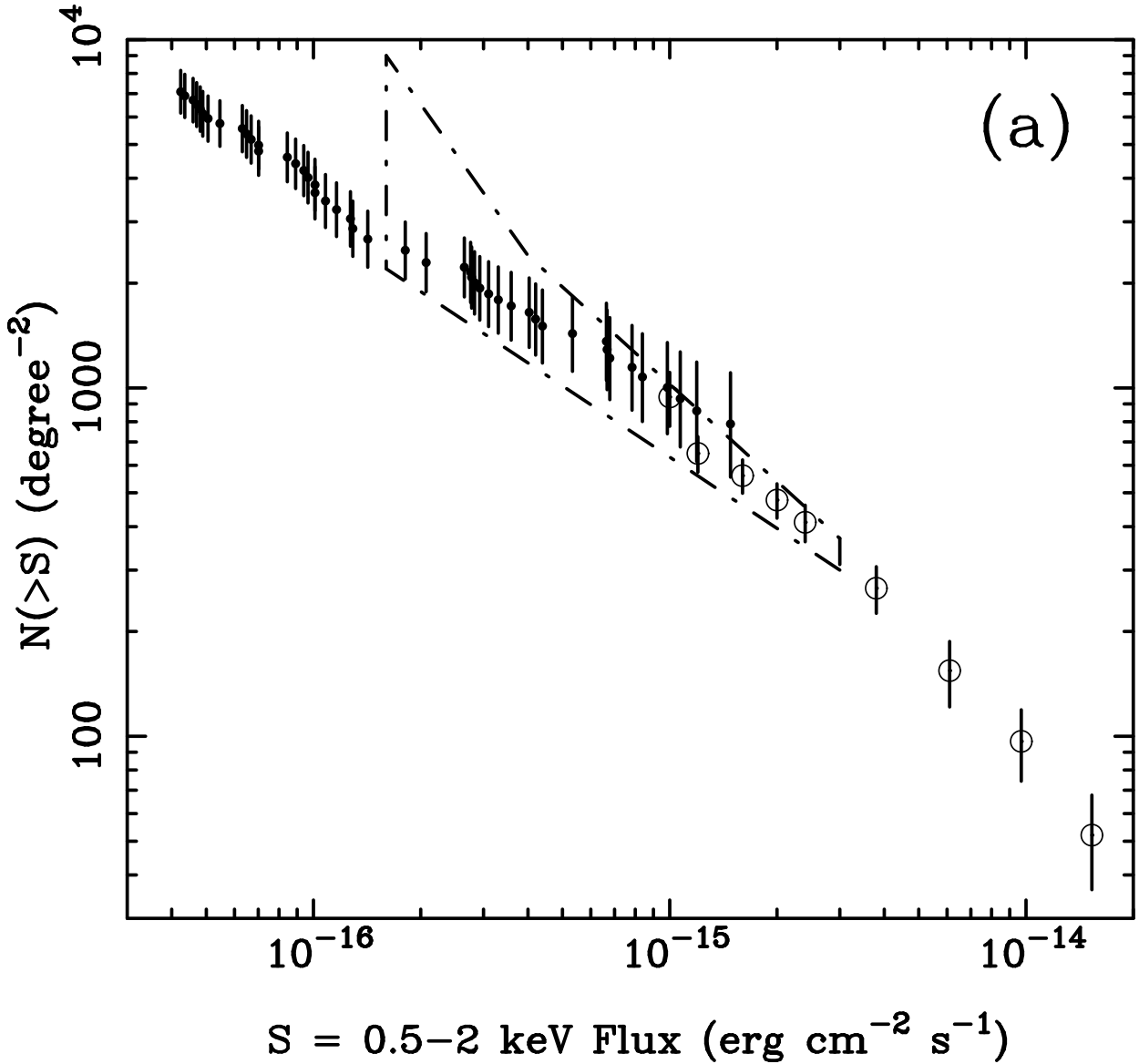}
\vspace{-0.1truein}
\caption{Number of sources, $N(>S)$, brighter 
than a given flux, $S$, for the (a) soft 
band and (b) hard band. In panel (a) the solid dots are our data points, the 
open circles are the \rosat\ Lockman Hole data points from 
Hasinger et~al. (1998), and the dot-dashed ``bow-tie'' region shows the 
\rosat\ Lockman Hole fluctuation analysis results of Hasinger et~al. (1993). 
In panel (b) the solid dots are our data points, the open circles are the 
\asca\ Large Sky Survey and Deep Sky Survey data points from 
Ueda et~al. (1998) and Ogasaka et~al. (1998), and the 
dot-dashed bow-tie region shows the multiple-field fluctuation analysis 
results of Gendreau, Barcons, \& Fabian (1998). The Ueda et~al. (1998),
Ogasaka et~al. (1998), and 
Gendreau et~al. (1998) results have been corrected to the 2--8~keV
band assuming a $\Gamma=1.4$ power-law spectrum. The error bars on our
data points have been computed following Gehrels (1986) for $1\sigma$.
We have only plotted data points where more than 10 sources were used
to determine the number counts.}
\end{figure}

\begin{figure}
\epsscale{1.0}
\figurenum{11e}
\plotone{brandt.fig13b.ps}
\end{figure}


\begin{figure}
\epsscale{1.0}
\figurenum{14}
\plotone{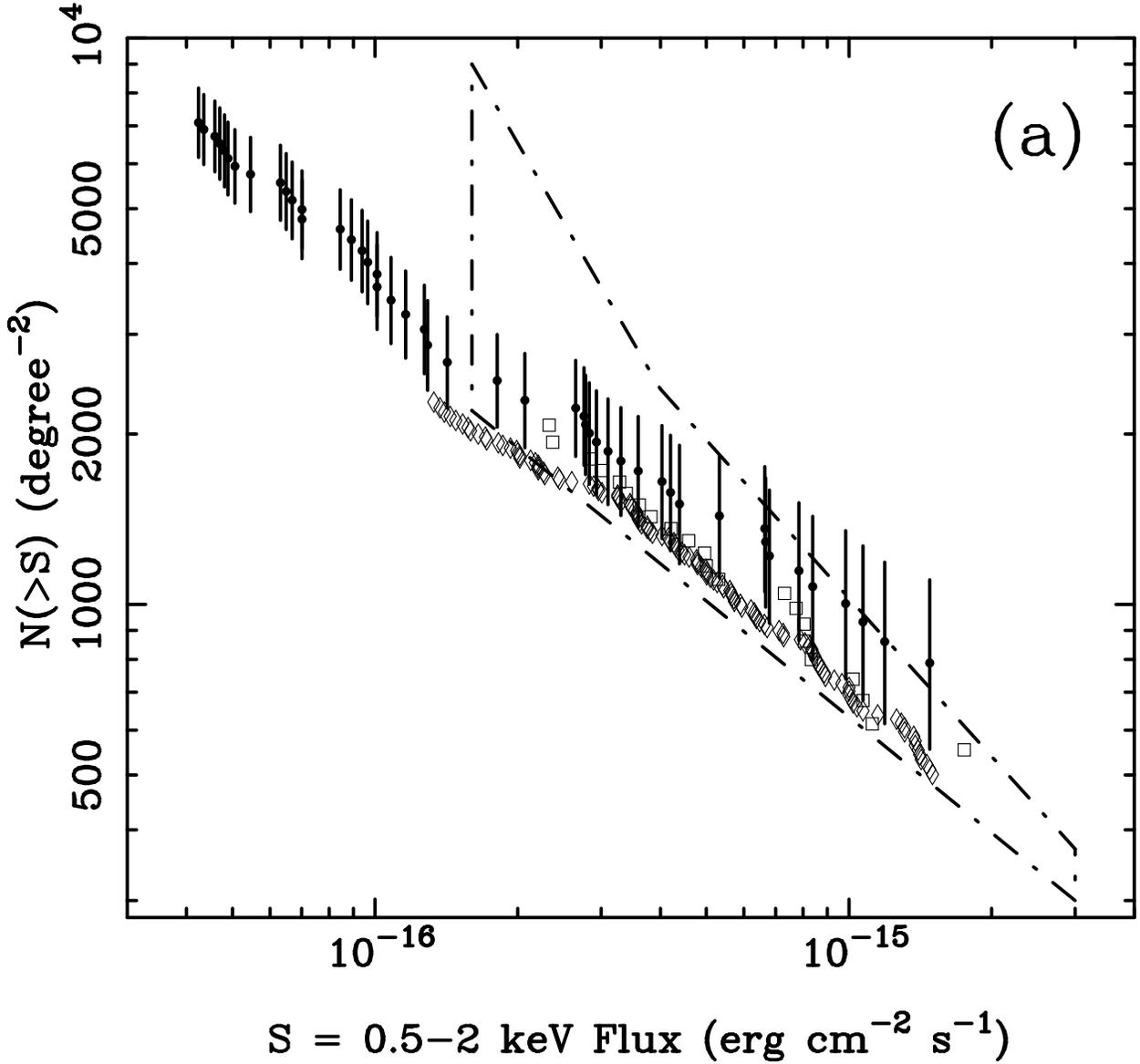}
\vspace{-0.1truein}
\caption{Comparisons of our $N(>S)$ curves (solid dots) with those from 
Mushotzky et~al. (2000; open squares) and
Tozzi et~al. (2001; open diamonds) 
for the (a) soft band and (b) hard band. 
The dot-dashed regions show the fluctuation analysis results
described in Figure~13. In panel (b) the 
Mushotzky et~al. (2000) and Tozzi et~al. (2001) 
results have been corrected to the 2--8~keV band 
assuming a $\Gamma=1.4$ power-law spectrum.}
\end{figure}

\begin{figure}
\epsscale{1.0}
\figurenum{11e}
\plotone{brandt.fig14b.ps}
\end{figure}


\begin{figure}
\epsscale{1.0}
\figurenum{15}
\plottwo{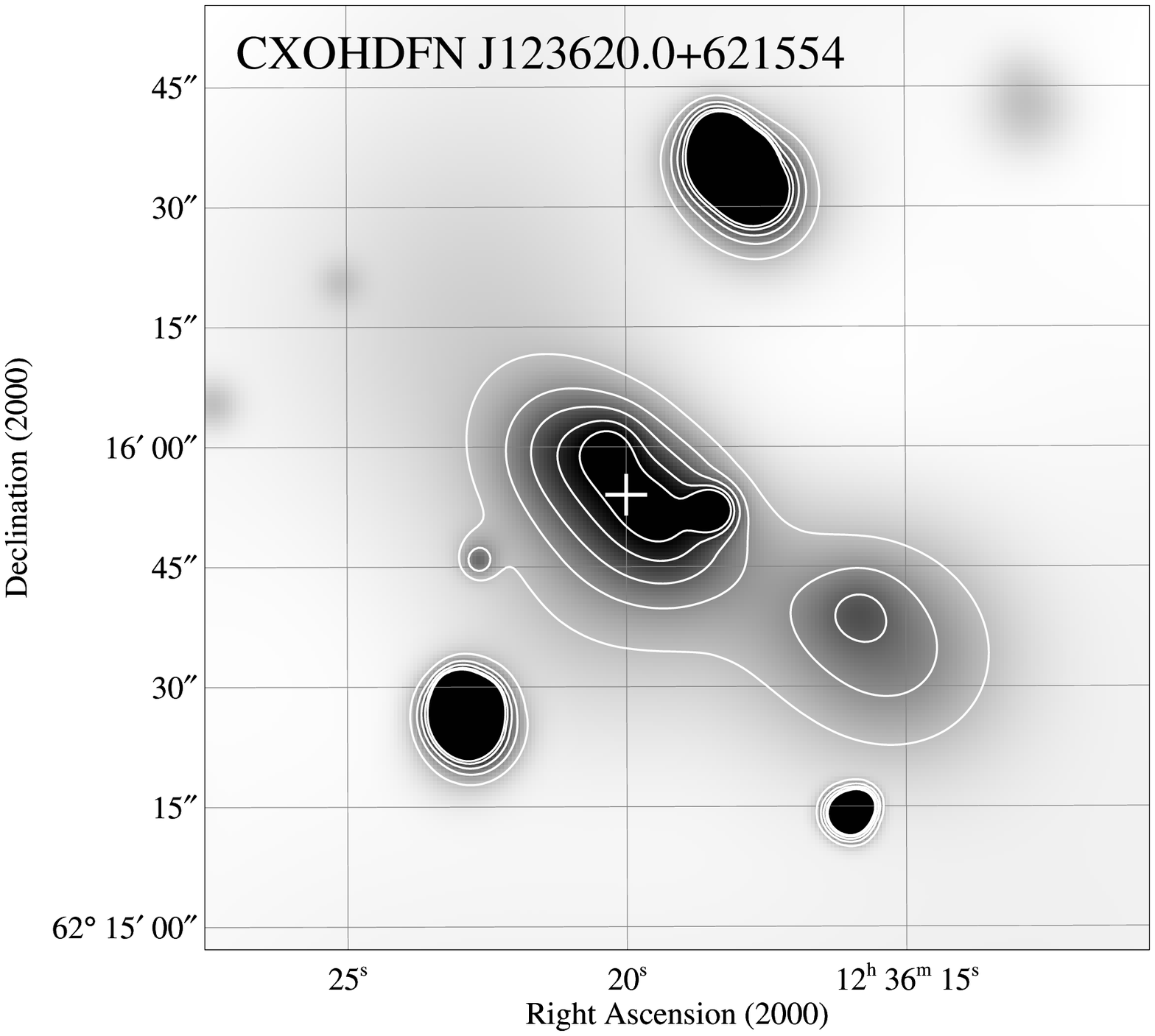}{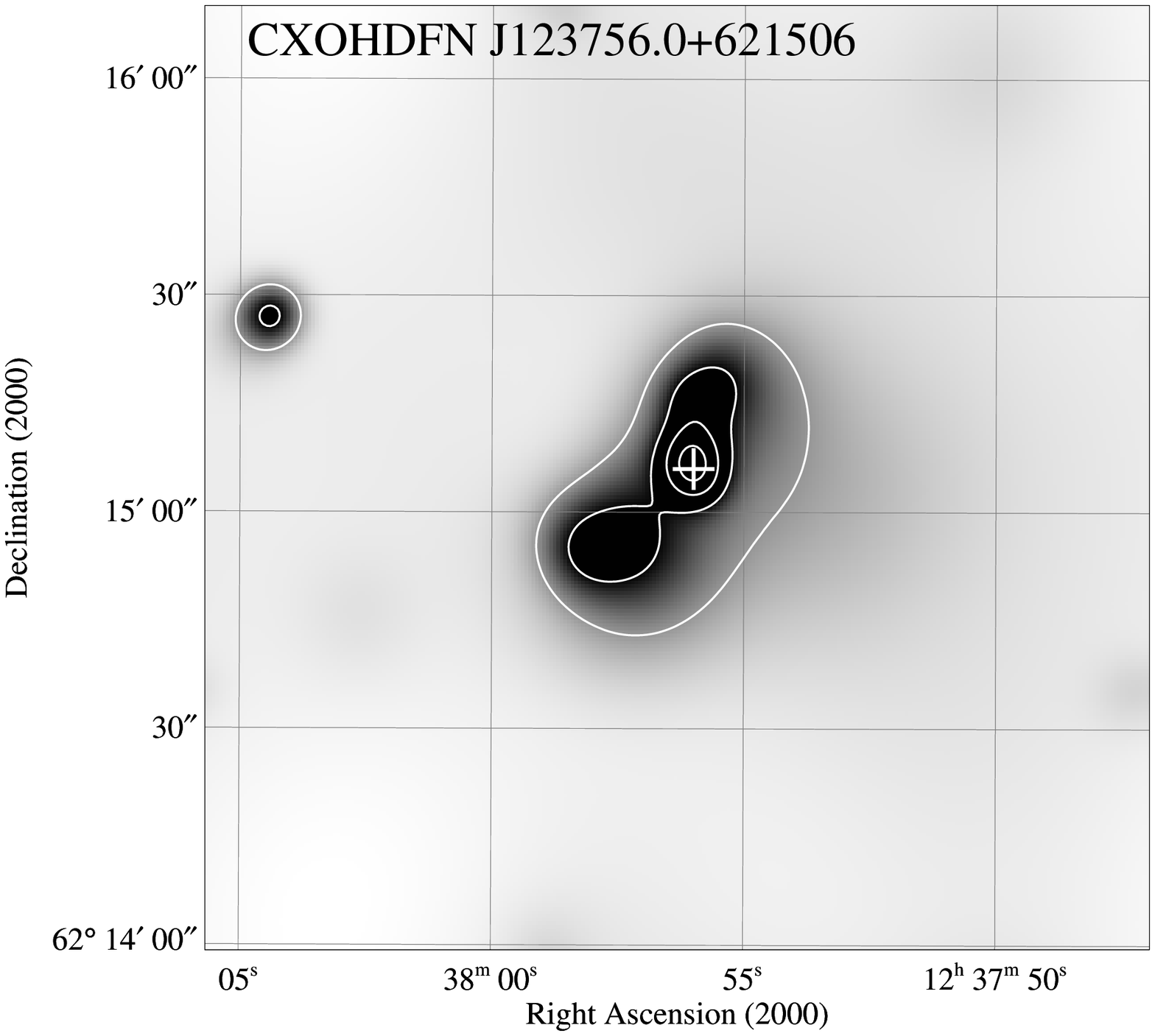}
\vspace{-0.1truein}
\caption{Adaptively smoothed and contoured \chandra\ soft-band images of the spatially
extended X-ray sources
CXOHDFN~J123620.0+621554 (contours are at 55, 65, 75, 85, and 95\%\ of the maximum pixel value) and 
CXOHDFN~J123756.0+621506 (contours are at 30, 50, 70, and 90\% of the maximum pixel value). 
The adaptive smoothing has been performed using the code of 
Ebeling et~al. (2001) at the $2.5\sigma$ level,
and the grayscales are linear. A cross indicates the adopted ``central'' 
position of each extended source. The point sources in each image 
can be used to judge the appropriate PSF sizes.}
\end{figure}


\begin{figure}
\epsscale{0.85}
\figurenum{16}
\plotone{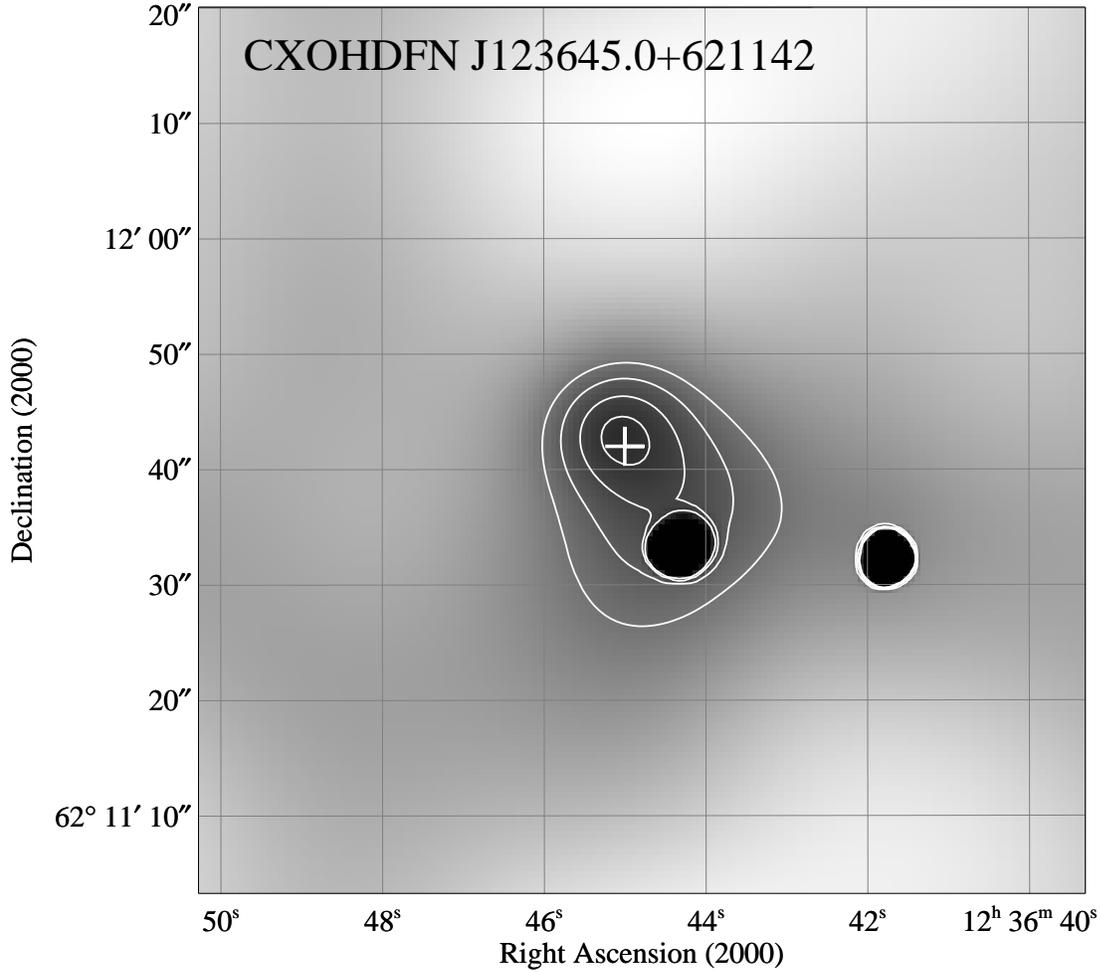}
\vspace{-0.1truein}
\caption{Adaptively smoothed and contoured \chandra\ soft-band image of the probable 
spatially extended X-ray source
CXOHDFN~J123645.0+621142 (contours are at 85, 90, 95, and 99\%\ of the maximum pixel value). 
The adaptive smoothing has been performed using the code of 
Ebeling et~al. (2001) at the $2.5\sigma$ level,
and the grayscales are linear. A cross indicates the adopted ``central'' 
position of the extended source. The point sources in the image 
can be used to judge the appropriate PSF size. 
The point source within the diffuse emission, CXOHDFN~J123644.3+621132,  
is associated with the $z=1.050$ FR~I radio galaxy VLA~J123644.3+621133.}
\end{figure}


\end{document}